\documentclass{article}
\usepackage{arxiv}

\usepackage{hyperref}
\usepackage{microtype}

\usepackage{amssymb}
\usepackage{amsmath}
\usepackage{amsthm}
\usepackage{mathdots}
\usepackage{mathtools}

\usepackage{tensor}

\usepackage{urwchancal}
\DeclareFontFamily{OT1}{pzc}{}
\DeclareFontShape{OT1}{pzc}{m}{it}{<-> s * [1.10] pzcmi7t}{}
\DeclareMathAlphabet{\mathpzc}{OT1}{pzc}{m}{it}

\usepackage{graphicx}
\usepackage{ifpdf}
\ifpdf
\usepackage{epstopdf}
\PrependGraphicsExtensions{.svg}
\fi

\usepackage{xypic}

\providecommand{\R}{\mathbb{R}}

\providecommand{\SO}{\mathbf{SO}}
\providecommand{\SL}{\mathbf{SL}}

\providecommand{\SE}{\mathbf{SE}}

\providecommand{\MR}{\mathbf{MR}}

\providecommand{\grpG}{\mathbf{G}}

\providecommand{\grpR}{\mathbf{R}}

\providecommand{\gothg}{\mathfrak{g}}

\providecommand{\gothm}{\mathfrak{m}}

\providecommand{\gothr}{\mathfrak{r}}

\providecommand{\gothX}{\mathfrak{X}} 

\providecommand{\so}{\mathfrak{so}}

\providecommand{\Sph}{\mathrm{S}}

\providecommand{\calM}{\mathcal{M}}
\providecommand{\calN}{\mathcal{N}}

\providecommand{\vecL}{\mathbb{L}}

\providecommand{\PD}{\mathbb{S}_+} 

\providecommand{\tT}{\mathrm{T}} 

\providecommand{\Id}{I} 

\providecommand{\eb}{\mathbf{e}} 

\providecommand{\GP}{\mathbf{N}} 
\providecommand{\WP}{\mathbf{W}} 

\providecommand{\vmu}{\mu_{v}} 

\providecommand{\ynu}{\nu_{y}} 

\DeclareMathOperator{\spn}{span}

\DeclareMathOperator{\stab}{stab}
\DeclareMathOperator{\Ad}{Ad}
\DeclareMathOperator{\ad}{ad}

\DeclareMathOperator{\image}{im}

\providecommand{\PT}{\mathbf{P}}

\providecommand{\id}{\mathrm{id}} 

\providecommand{\Lyap}{\mathcal{L}} 

\providecommand{\td}{\mathrm{d}}
\providecommand{\tD}{\mathrm{D}}

\providecommand{\ddt}{\frac{\td}{\td t}}

\providecommand{\dt}{\td t}

\providecommand{\mr}[1]{\mathring{#1}} 

\usepackage{accents}
\makeatletter
\providecommand{\scirc}{%
    \hbox{\fontfamily{\rmdefault}\fontsize{0.4\dimexpr(\f@size pt)}{0}\selectfont{\raisebox{-0.52ex}[0ex][-0.52ex]{$\circ$}}}}

\makeatother

\mathchardef\mhyphen="2D

\providecommand{\etal}{\textit{et al.}~}

\newcommand{\papercitation}{
R.~Mahony, P.~van Goor, and T.~Hamel, ``Observer design for nonlinear systems with equivariance,'' \emph{Annual Review of Control, Robotics, and Autonomous Systems}, vol.~5, no.~1, pp.~221--252, 2022. (DOI: \DOInumber) URL: \linktopaper}
\newcommand{\linktopaper}{https://doi.org/10.1146/annurev-control-061520-010324}
\newcommand{\DOInumber}{10.1146/annurev-control-061520-010324}
\newcommand{\publicationdetails}
{
\papercitation \\
}
\newcommand{\publicationversion}
{Author accepted version}

\begin{document}

\title{Observer Design for Nonlinear Systems with Equivariance}
\headertitle{Equivariant Observers}

\author{
\href{https://orcid.org/0000-0002-7803-2868}{\includegraphics[scale=0.06]{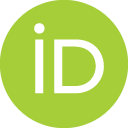}\hspace{1mm}
Robert Mahony}
\\
	Systems Theory and Robotics Group \\
	Australian National University \\
    ACT, 2601, Australia \\
	\texttt{Robert.Mahony@anu.edu.au} \\
	\And	\href{https://orcid.org/0000-0003-4391-7014}{\includegraphics[scale=0.06]{orcid.png}\hspace{1mm}
Pieter van Goor}
\\
	Systems Theory and Robotics Group \\
	Australian National University \\
    ACT, 2601, Australia \\
	\texttt{Pieter.vanGoor@anu.edu.au} \\
	\And	\href{https://orcid.org/0000-0002-7779-1264}{\includegraphics[scale=0.06]{orcid.png}\hspace{1mm}
Tarek Hamel}
\\
I3S-CNRS, 2000 route des Lucioles, \\
University C\^ote d'Azur and Insitut Universitaire de France, \\
06903 Sophia-Antipoles cedex, France.
	\texttt{thamel@i3s.unice.fr} \\
}

\maketitle

\begin{abstract}
Equivariance is a common and natural property of many nonlinear control systems, especially those associated with models of mechatronic and navigation systems.
Such systems admit a symmetry, associated with the equivariance, that provides structure enabling the design of robust and high performance observers.
A key insight is to pose the observer state to lie in the symmetry group rather than on the system state space.
This allows one to define a globally defined intrinsic equivariant error but poses a challenge in defining internal dynamics for the observer.
By choosing an equivariant lift of the system dynamics for the observer internal model we show that the error dynamics have a particularly nice form.
Applying the methodology of Extended Kalman Filtering (EKF) to the equivariant error state yields the Equivariant Filter (EqF).
The geometry of the state-space manifold appears naturally as a curvature modification to the classical EKF Riccati equation.
The equivariant filter exploits the symmetry and respects the geometry of an equivariant system model and yields high performance robust filters for a wide range of mechatronic and navigation systems.
\end{abstract}



\section{Introduction}\label{sec:intro}

The celebrated scientist Stanislaw Ulam once stated that ``studying nonlinear systems is like [...] studying non-elephant animals''  \cite{1987_campbell_ParadigmsPracticalities}.
There is such a vast array of behaviours and structures encapsulated in the term \emph{nonlinear} that any analysis can only apply to a subclass of possible systems.
Real world observer problems, even when highly non-linear, often have structure that can be exploited in the design of algorithms.
For mechatronic and navigation systems one of the most common structures is symmetry.
The present paper concerns how to exploit such symmetry to design high performance observers for nonlinear control systems.

Formally, symmetry is a property of the defining equations of motion of a system.
A system is said to by symmetric if there is a transformation of the state-space (the symmetry) for which the equations of motion of the system are unchanged (an invariance) or change in a structured manner (an equivariance).
For example, consider directional kinematics on a sphere
\begin{align}
\dot{\eta} = - \Omega \times \eta
\label{eq:direction_kinematics}
\end{align}
where the state $\eta \in \Sph^2$ is a direction on a sphere, $\Omega \in \R^3$ is an input, and $\times$ is the usual vector cross product.
For an observer problem, measurements of the inputs $\Omega$ are provided by a velocity sensor such as an inertial measurement unit with rate gyroscopes providing velocity measurements in the body-fixed frame \cite{2020_Mahony_EquivariantSystemsTheory}.
The goal will be to estimate the direction $\eta$ from (noisy) velocity measurements and some additional (noisy) direction measurements.
The directional kinematics \eqref{eq:direction_kinematics} are \emph{equivariant} under multiplication $\eta \mapsto Q^\top \eta$ by a (constant) rotation matrix $Q \in \SO(3)$.
That is
\[
\ddt (Q^\top \eta) = Q^\top \dot{\eta} = Q^\top (\Omega \times \eta) = (Q^\top \Omega) \times (Q^\top \eta).
\]
The transformed kinematics are written in terms of a transformed state variable $\eta' := Q^\top \eta$ and a transformed input variable $\Omega' := (Q^\top \Omega)$.
The symmetry of the equations of motion is expressed in the sense that the  new kinematics $\dot{\eta}' = -\Omega' \times \eta'$ are identical to the original equations.
The input transformation $\Omega' = Q^\top \Omega$ is important, it encodes the \emph{equivariance} of the system.
\emph{Invariant} systems are a subclass of equivariant systems for which the input transformation is trivial, that is the transformed and original system use the same input.
Invariance is particularly relevant for autonomous dynamical systems (with no inputs) and is a foundation principle of classical and modern mechanics.
The presence of inputs leads naturally to equivariance playing an analogous role for nonlinear control systems.

In this paper, we consider nonlinear control systems with symmetry and propose an approach for the design of nonlinear observers that is based on two simple precepts:
\begin{itemize}
  \item Exploit the \emph{symmetry} of a system.
  \item Respect the \emph{geometry} of a system.
\end{itemize}

The first precept leads to a foundation principle of equivariant observer design: \emph{that the observer state is formulated on the symmetry group rather than the state-space}.
For example, in the direction estimation problem discussed earlier, the symmetries are the set of all rotations in the special orthogonal group $\SO(3)$.
The equivariant design approach poses the observer state $\hat{Q} \in \SO(3)$ as a rotation matrix while the estimate of the state is recovered by applying the symmetry $\hat{\eta} := \hat{Q}^\top \mr{\eta}$ to an arbitrary constant \emph{origin} direction $\mr{\eta} \in \Sph^2$.
Note that the observer-state lies in a completely different, and higher dimensional, manifold $\hat{Q} \in \SO(3)$ compared to the system-state $\eta \in \Sph^2$, a general characteristic of equivariant observer design.
This simple reparametrization is more powerful than it may appear at first glance.
Whereas $\hat{\eta}$ is just a point on the sphere, the equivariant observer state $\hat{Q}$ corresponds to a symmetry diffeomorphism of the \emph{whole} space $\hat{Q}^\top : \Sph^2 \to \Sph^2$.
Applying (the inverse of) this symmetry to the true state defines the \emph{equivariant error} $e = \hat{Q} \eta \in \Sph^2$; a globally defined intrinsic error for the observer design problem.
This construction should be compared to a classical observer design on a manifold where the observer/state error depends on local or embedded coordinates and is neither global nor intrinsic.
The goal of equivariant observer design is to find dynamics on the symmetry group (the set of rotations $\SO(3)$ in this case) such that the observer state (the rotation $\hat{Q}$ in this case) evolves to force the error $e \to \mr{\eta}$ to converge to the known fixed origin, leading to the state estimate $\hat{\eta} \to \eta$ to converge to the true-state.

Posing the observer-state as an element of the symmetry group raises the question of how to define observer dynamics, since the obvious copy of the state dynamics as an internal model is no longer an option.
Recent work \cite{2020_Mahony_EquivariantSystemsTheory} showed that (for an equivariant system) there is always an \emph{equivariant lift} $\Lambda$ of the system dynamics.
The lift function defines a set of dynamics on the symmetry group that project down to the true system dynamics on the state-manifold and provide the structure of an internal model for the observer architecture.
For the simple example we are discussing\footnote{
Here $\Omega^\times$ denotes the $3 \times 3$ skew-symmetric matrix that encodes the vector cross product $\Omega^\times \eta = \Omega \times \eta$.
}
$\Lambda = \Omega^\times$ and the observer is
\[
\ddt \hat{Q} = \hat{Q} \Omega^\times + \Delta \hat{Q}
\]
where the observer depends on an additional correction term $\Delta_t \in \so(3)$, a skew-symmetric matrix in the Lie-algebra of $\SO(3)$ that is used to steer $e \to \mr{\eta}$.
For $\Delta_t \equiv 0$ the solutions of these observer-dynamics project $\hat{\eta}(t) = \hat{Q}(t)^\top \mr{\eta}$ to solutions of the system, encoding the internal model principle fundamental in observer design despite the difference in observer-state and system-state spaces.

It remains to ``design'' the correction term $\Delta_t$.
Here the equivariant properties of the lift $\Lambda$ play a key role in ensuring that the error dynamics are both globally defined and locally well conditioned.
Indeed, for the simple direction estimation example
\begin{align*}
\ddt e & = \ddt \hat{Q} \eta
= \hat{Q} \Omega^\times \eta + \Delta_t \hat{Q} \eta - \hat{Q} (\Omega \times \eta)
 =  \Delta_t e.
\end{align*}
Although, the velocity dependent part of the error dynamics for a general system do not cancel, they can always be written as a globally defined function of the equivariant error $e$ and a known \emph{origin} input.
This structure in the error dynamics provides a foundation for a whole range of global nonlinear observer design methodologies
\cite{
RM_2011_Hua_cdc,
2010_vasconcelos_NonlinearPositionAttitude,
2011_madgwick_EfficientOrientationFilter,
2011_Hamel_cdc,
2012_Trumpf_TAC,
2012_grip_AttitudeEstimationUsing,
2012_batista_SensorBasedGloballyAsymptotically,
2014_izadi_RigidBodyAttitude,
2015_Hua,
2016_allibert_VelocityAidedAttitude,
2016_hua_StabilityAnalysisVelocityaided,
2017_berkane_HybridGlobalExponential,
2017_LeBras,
2015_Hua,
2018_Zlotnik_GradientSLAM,
MiaTayeb2019,
2019_hua_FeaturebasedRecursiveObserver,
2020_Hua_Automatica}
as well as opening the door for principled observer design based on linearisation
\cite{
Bonnabel2006_acc,Martin2007,
2008_martin_InvariantObserverEarthVelocityAided,
2008_Bonnabel_TAC,
bonnabel2009,
2009_bonnabel_InvariantExtendedKalman,
RM_2012_Zamani_TAC,
bourmaud2013_discreteEKF,
2015_bourmaud_ContinuousDiscreteExtendedKalman,
2015_barrau_IntrinsicFilteringLie,
2016_Saccon_TRO,
2017_Barrau_tac,
2018_barrau_InvariantKalmanFiltering,
2019_Lavoie_InvariantHinf,
2020_Phogat_iEKF,
vanGoor2020_EqF.cdc,
vanGoor2020_EqF.tac}.

\begin{figure}
  \includegraphics[width=0.8\linewidth]{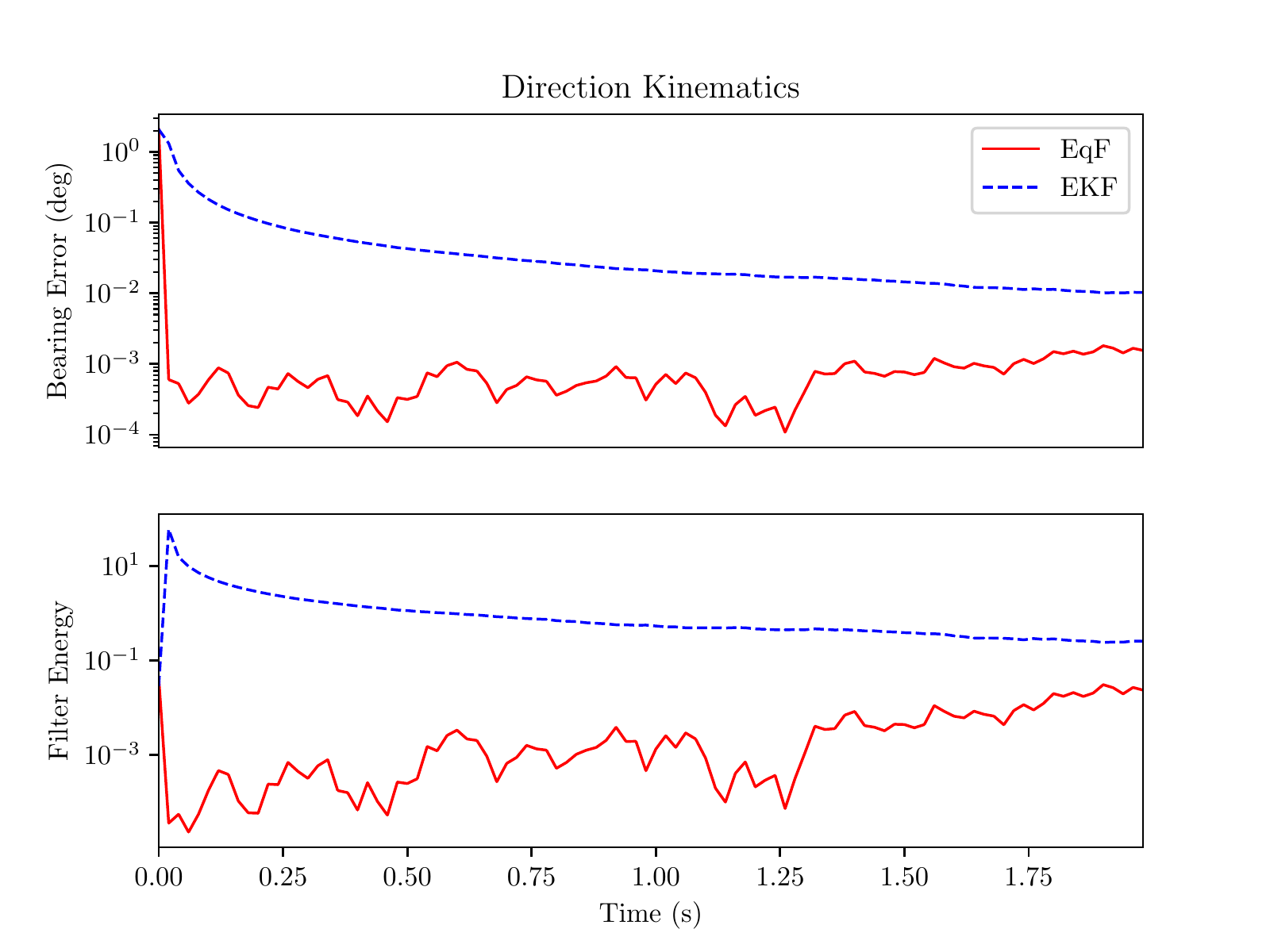}
  \caption{Direction estimation on the sphere.
  Comparison of an EqF in normal coordinates (solid red line) versus an EKF in stereographic coordinates (dashed blue line).
  }
  \label{fig:eqf_sphere}
\end{figure}

In the present paper, we provide a principled analysis of a linearising observer design by adapting the extended Kalman filter to equivariant error coordinates \cite{vanGoor2020_EqF.cdc,vanGoor2020_EqF.tac}.
Here we draw on the second precept \emph{to respect the geometry}.
Only a single set of local coordinates, centred on the origin in error coordinates, is required in contrast to a classical EKF that depends on linearisations taken along a time-varying trajectories, and this leads to improved filter performance.
Moreover, as shown in the seminal work of Barrau and Bonnabel \cite{2017_Barrau_tac}, there is a canonical choice of local coordinates (the normal coordinates) for which the linearisation error is minimized, leading to improved filter performance.
Finally, a careful application of the principles of extended Kalman filtering in error coordinates, taking account of the geometry of the space, leads naturally to a curvature modification of the Riccati equation corresponding to a parallel transport of the covariance estimate.
We show that including this term in the Ricatti also improves the filter response, particularly during the transient phase.
The resulting filter is termed the \emph{Equivariant Filter (EqF)} (see also \cite{vanGoor2020_EqF.cdc,vanGoor2020_EqF.tac}) and the design approach provides high performance filters for a wide range of equivariant nonlinear systems.
Even for the simple direction estimation example discussed above, for which a classical EKF using stereographic coordinates is highly effective, the performance gain in following principled equivariant observer design is clearly shown in Figure~\ref{fig:eqf_sphere} (see Section \ref{sec:simulations} for details of the simulation).

\section{Literature Review}\label{sec:lit_review}

Symmetry in mechatronic systems is a natural consequence of the invariance of the physical laws that govern their dynamics.
The field of geometric control systems \cite{Bullo2005_GeometricControlBook,Jurdjevic1997_GeometricControlTheory} is founded on this perspective and was highly active from the seventies through to the nineties with work still ongoing.
Although early work in this field \cite{1981_VdSchaft_SymmetryHamilton,1985_Grizzle_StructureNLSym,1985_Nijmeijer_PartialSyms} considered symmetry for general control system models, the field matured towards using the Euler-Lagrange and Hamiltonian frameworks and exploiting symmetry for structural analysis of these dynamics \cite{Bullo2005_GeometricControlBook,Jurdjevic1997_GeometricControlTheory}.
Despite some effort  \cite{Aghannan2003,2004_Maithripala_acc,2010_bonnabel_SimpleIntrinsicReducedObserver}
this body of theory has not translated easily into observer and filter design.
In contrast, working with simpler kinematic models has had significant impact in navigation systems \cite{2018_barrau_InvariantKalmanFiltering}.
An early contribution in this direction was the work of Salucedean \cite{Sal1991_TAC} on attitude estimation of a satellite.
The original theory was extended a decade later \cite{VikFos_2001_CDC,ThiSan2003_TAC} to incorporate bias estimation in the sensor measurements.
This work was done in parallel with the established Multiplicative Extended Kalman Filter (MEKF) attitude estimators that had been first developed in the eighties \cite{1982_lefferts_KalmanFilteringSpacecraft} but were still under  active research at the turn of the century \cite{2003_markley_AttitudeErrorRepresentations,2006_choukroun_NovelQuaternionKalman,2007_Crassidis.JGCD}.
It was around 2005 that the importance of robust attitude filtering for the growing Unmanned Aerial Vehicle (UAV) industry became clear and the resulting interest lead to a burst of activity
\cite{2005_MahHamPfl-C64,Bonnabel2005_ch,2006_HamMah_ICRA,Bonnabel2006_acc,Martin2007} leading to the seminal papers \cite{2008_MahHamPfl.TAC,bonnabel2009} proposing simple, robust observers that were an enabling technology in the development of commercial systems.
Although the quality of Inertial Measurements Units (IMUs) has now improved, these early attitude observers were key in overcoming the very high noise levels and unreliability of the IMUs available to the early UAV systems.
There was a significant body of work done in the 2010s that extended and developed these ideas for attitude estimation \cite{2011_madgwick_EfficientOrientationFilter,2012_Trumpf_TAC,2012_grip_AttitudeEstimationUsing,2012_batista_SensorBasedGloballyAsymptotically,2014_izadi_RigidBodyAttitude,2017_berkane_HybridGlobalExponential}.
Analogous approaches have considered the Special Euclidean group $\SE(3)$ for pose estimation \cite{2010_vasconcelos_NonlinearPositionAttitude,RM_2011_Hua_cdc,2015_Hua,MiaTayeb2019,2017_LeBras}, the Special Linear group $\SL(3)$ for homography estimation
\cite{2011_Hamel_cdc,2015_Hua,2019_hua_FeaturebasedRecursiveObserver,2020_Hua_Automatica},
and a number of works on velocity aided attitude estimation \cite{Bonnabel2006_acc,2008_martin_InvariantObserverEarthVelocityAided,2016_allibert_VelocityAidedAttitude,2016_hua_StabilityAnalysisVelocityaided}.
A key contribution here was in the work of Bonnabel \etal \cite{2008_Bonnabel_TAC,2009_bonnabel_InvariantExtendedKalman} who introduced a new group structure, later denoted by $\SE_m(n)$ \cite{2016_Barrau_arxive}, that allowed second order translation kinematics to be modelled in the same group structure as first order kinematics in attitude.
This group structure was also exploited to model the Simultaneous Localisation and Mapping problem \cite{2016_Barrau_arxive,2017_zhang_ConvergenceConsistencyAnalysis,2017_Mahony_cdc,2021_mahony_HomogeneousSpaceGeometry}. In parallel Zlotnik \etal used a direct product structure \cite{2018_Zlotnik_GradientSLAM} for the SLAM problem.

There are several works over the last fifteen years that have developed general theory for classes of systems
\cite{bonnabel2009,RM_2013_Mahony_nolcos,2015_bourmaud_ContinuousDiscreteExtendedKalman,2017_Barrau_tac,2020_Mahony_EquivariantSystemsTheory,2020_Ng_cdc}.
Bonnabel and coauthors have built a theory for the Invariant Extended Kalman Filter (IEKF) in a series of papers \cite{2007_bonnabel_LeftinvariantExtendedKalman,2009_bonnabel_InvariantExtendedKalman,2015_barrau_IntrinsicFilteringLie}.
Although similar in spirit to the multiplicative extended Kalman filter \cite{1982_lefferts_KalmanFilteringSpacecraft,2007_Crassidis.JGCD}
and the error state Kalman filter \cite{1999_roumeliotis_CircumventingDynamicModeling,2017_sola_QuaternionKinematicsErrorstate}, the IEKF provides a clear design methodology for general systems posed on Lie-groups.
This perspective has provided significantly more insight into the structural properties of observer systems on Lie-groups and lead to the definition of the ``group affine'' property  \cite{2016_Barrau_arxive,2017_Barrau_tac,2020_Mahony_mtns} for which it is possible to prove powerful stability results for highly nonlinear systems.
Stability and robustness of high performance filters can also be guaranteed by the direct design of Ricatti observers \cite{2018_Hamel_TAC,MiaBerTay2021}.
An advantage of this approach is that the filters are not tied to a fixed linearisation structure, however, there is no standardised design methodology and each example must be considered independently.
The majority of approaches in the literature, including IEKF, MEKF, and the Riccati observers, are only applicable to systems posed directly on a Lie-group.
Equivariant systems on smooth homogeneous spaces, that is, systems on manifolds that support a smooth group action from a higher dimensional Lie-group \cite{RM_2013_Mahony_nolcos}, is a more general class of systems to which analogous observer design principles can be applied.
This perspective goes back to the origins of modern equivariant observer theory where direction estimation on $\Sph^2$, an homogeneous space under rotation action by $\SO(3)$, was proposed in \cite{Metni2005} and lifted to $\SO(3)$ \cite{2008_MahHamPfl.TAC} to obtain the complementary filter for attitude discussed earlier.
A similar perspective also led from \cite{2007_Bonnabel.thesis} to \cite{bonnabel2009} and recent work on Ricatti observers for direction estimation \cite{hua2018}.
Unfortunately, direction kinematics was the only known example of a robotics application involving a true homogeneous system structure until the SLAM problem was considered over ten years later \cite{2017_Mahony_cdc,2021_mahony_HomogeneousSpaceGeometry}.
The gauge invariance inherent in the SLAM problem induces a homogeneous space structure that underlies recent work by the authors \cite{vanGoor2020_EqF.cdc,2020_vanGoor_ifac,2019_vanGoor_cdc,2021vangoorEquivariantFilter_ICRA}.
This perspective is critically important for the visual SLAM problem where cameras are the primary exteroceptive sensor and the SLAM problem can no longer be modelled \cite{vanGoor2020_EqF.tac} using the $\SE_{n+1}(3)$ geometry introduced by Barrau \etal \cite{2016_Barrau_arxive}.
In other recent work, Joshi \etal has also considered more general bundle structures for the equivariant observer problem \cite{joshi2020bundle},
Izadi \etal has exploited variational approaches \cite{2014_izadi_RigidBodyAttitude}, Lavoie considered $H_\infty$ observers \cite{2019_Lavoie_InvariantHinf}, Bourmaud \etal \cite{bourmaud2013_discreteEKF,2015_bourmaud_ContinuousDiscreteExtendedKalman} proposed extended Kalman and unscented filters on Lie groups from first principles, Ng \etal introduced a semi-direct group structure  \cite{2019_Ng_cdc,2020_Ng_cdc} for second order kinematics, while Phogat \etal considered a direct product structure for general second order systems on Matrix Lie groups  \cite{2020_Phogat_iEKF}.

In order to improve filter performance above the second order optimality in local coordinates provided by the extended Kalman filter,
Bonnabel and coauthors have proposed the invariant unscented Kalman filter
\cite{2017_brossard_UnscentedKalmanFilteringb,2018_brossard_UnscentedKalmanFiltera,2020_brossard_CodeUnscentedKalmanb} that exploits the invariant structure of the problem to transport the sample points of the unscented filter.
A similar algorithm was proposed in Bourmaud \etal \cite{2015_bourmaud_ContinuousDiscreteExtendedKalman}.
This approach can be generalised to any parallelizable manifold \cite{2020_brossard_CodeUnscentedKalmanb} and corresponds to placing a flat
geometric structure on the space.
Using a flat geometry goes counter to other recent work Loianno \etal \cite{2016_loianno_VisualInertialOdometry} in applying the UKF on $\SE(3)$.
Curvature also occurs naturally in the Geometric Approximate Minimum Energy (GAME) filters \cite{RM_2012_Zamani_TAC,2016_Saccon_TRO,Berger_2015_SecondOrderSE3,Berger_2017_SecondOrderFilterSE3}.
Second order filters \cite{1982_Maybeck_StochasticModels_v2} have been considered for some time in the INS field \cite{2003_markley_AttitudeErrorRepresentations,2004_Markley_AttitudeQuaternion}
and these terms turn out to be the same as the curvature terms found in GAME filters on $\tT \SO(3)$ \cite{RM_2012_Zamani_TAC,2016_Saccon_TRO}.
Questions around curvature and its connection with choice of local coordinates, linearisation error, second order filter terms, unscented filter design and filter performance are current research topics in equivariant

\section{Systems with Symmetry}\label{sec:symmetry}

In this section we discuss the formal structure of symmetry for the class of
affine control systems.
The concepts of invariance and equivariance are formulated in a unified manner as a symmetry property of the affine subspace of vector fields associated with the system.
This is an elegant and powerful way to understand the structure of symmetry for a control system with inputs.
From there we introduce the concept of lift and show how this leads to a system defined on the symmetry group that will be used as the internal model for the observer design in Section \ref{sec:Observer}.
The lift must satisfy two properties, pre-image and equivariance, that ensure firstly that it is a true pre-image of the system dynamics, and secondly that the associated error dynamics have nice symmetry properties.

Consider a system
\begin{subequations}\label{eq:system}
\begin{align}
\dot{\xi} & = f(\xi, u)  \label{eq:system_f}\\
y & = h (\xi) \label{eq:system_h}
\end{align}
\end{subequations}
for $\xi \in \calM$ the state-space, a smooth manifold, $y \in \calN$ the output-space, a smooth manifold, $u \in \vecL$ the input-space, a linear vector space.
The output function $h : \calM \to \calN$ is smooth and well defined everywhere.
The system function $f : \calM \times \vecL \to T \calM$ is smooth and affine in the second argument.
That is, in local coordinates it can be written
\[
f(\xi, u) =  \sum_{i = 1}^l f_i(\xi) u^i + f_0(\xi)
\]
for $u = (u^1, \ldots, u^l) \in \vecL$ known (measured) input signals where $f_i$ are termed the \emph{input vector fields} and $f_0$ is termed the \emph{drift field}.

For example, consider second order linear kinematics in $\R^3$
\begin{subequations}
\label{eq:lin_kin}
\begin{align}
\dot{p} & = v \\
\dot{v} & =  a
\end{align}
\end{subequations}
where $\xi = (p, v) \in \calM = \R^3 \times \R^3$ is the state and $u = a \in \R^3$ is the input.
The drift field is $f_{0}(\xi) = (v,0)$.
The input vector fields are $f_i = (0,\eb_i)$ for $i =1 \ldots, 3$ where $\eb_i$ are the coordinate unit vectors, that is, $(0, {a}) = \sum_{i = 1}^3 f_i(\xi) a^i$.

For the case of direction kinematics \eqref{eq:direction_kinematics} discussed in the introduction, $\xi = \eta \in \Sph^2$ and the drift field $f_0(\xi) = 0$ is trivial.
The input vector fields are
\[
f_1 (\eta) = -\eb_1 \times \eta, \quad f_2 (\eta) =- \eb_2 \times \eta,\quad
f_3 (\eta) = -\eb_3 \times\eta
\]
and $\dot{\eta} = \sum_{j = 1}^3 f_j(\eta) \Omega_j$.

The functions $f_0, f_1, \ldots, f_l$ define the structure of the system considered.
These objects are vector fields on $\calM$, written $f_i \in \gothX(\calM)$, that smoothly assign a vector direction $f_i(\xi) \in \tT_\xi \calM $ to each element of $\calM$.
The set $\gothX(\calM)$ of smooth vector fields on $\calM$ is itself an (infinite dimensional) vector space under point wise addition and scalar multiplication.
That is, for all $f_1, f_2 \in \gothX(\calM)$ and $\alpha_1, \alpha_2 \in \R$ then $\alpha_1 f_1 + \alpha_2 f_2 \in \gothX(\calM)$.
The system function $f(\xi, u)$ can be interpreted as an affine map $f : \vecL \to \gothX(\calM)$ where $u \mapsto f_{u}$ defined by
\begin{align}
f_{u}(\xi) := f(\xi, u) = \sum_{i = 1}^l f_i(\xi) u^i + f_0(\xi)
\end{align}
We will also need notation for just the input or controlled part of system function independent of the drift term
\[
f^{\text{ctl}}_{u} :=  \sum_{i = 1}^l f_i(\xi) u^i.
\]
Thus, $f_u (\xi) = f^{\text{ctl}}_{u}(\xi) + f_0(\xi)$.

The core structure that enables equivariant observer design is the existence of a \emph{transitive family symmetries} on the state space $\calM$.
A symmetry is just a diffeomorphism of the state space with certain properties with respect to the system equations that we discuss further below.
Let $\grpG$ denote an index set for the family of diffeomorphisms and let $\phi_X : \calM \to \calM$ denote the diffeomorphism indexed by the element $X \in \grpG$.
The requirement that the family of symmetries is transitive means that for any two points in the state space, $\xi_1, \xi_2 \in \calM$ there is a symmetry $\phi_X : \calM \to \calM$ from the family such that $\xi_2 = \phi_X(\xi_1)$.
This assumption is necessary for equivariant observer design and is one of the places where the theory varies from that established in the geometric control field  \cite{Bullo2005_GeometricControlBook,Jurdjevic1997_GeometricControlTheory}.

Two diffeomorphisms can be concatenated to make a new diffeomorphism $\phi_X \circ \phi_Y : \calM \to \calM$.
Similarly, the inverse of a diffeomorphism $\phi_X^{-1}$ is also a diffeomorphism of $\calM$.
If for all $X, Y \in \grpG$ there is a $Z \in \grpG$ such that $\phi_Z = \phi_X \circ \phi_Y : \calM \to \calM$ and an $X^{-1} \in \grpG$ such that $\phi_{X^{-1}} = \phi_X^{-1}$, then the family of diffeomorphisms has the natural structure of a group.
In this case, the object $\phi : \grpG \times \calM \to \calM$ defined by
\begin{align}
\phi(X,\xi) : = \phi_X (\xi)
\label{eq:phi_X}
\end{align}
is termed a group action.
Groups and group actions are the natural structure in which to understand symmetries of state space.
In this paper we will consider \emph{right} handed group actions where $\phi_X \circ \phi_Y = \phi_{YX}$ and $YX$ is the group multiplication.
This is the natural choice for most robotics applications, particularly those with body fixed-frame sensing systems.
The reader is warned that most physics and mathematics texts are written using left handed actions.
Theory developed for right-handed symmetries can be transferred directly to left-handed symmetries without loss of generality by redefining the group multiplication, although this process does lead to considerable notational and conceptual complexity and in the authors opinion it is better to restrict to a single choice of handedness as we do in this paper.

Consider the direction kinematics \eqref{eq:direction_kinematics}.
The set of diffeomorphisms considered are
\begin{align}
\phi_Q (\eta) :=  Q^\top \eta.
\label{eq:atthi}
\end{align}
where $Q \in \R^{3 \times 3}$ is a rotation matrix.
The associated group is the special orthogonal group $\SO(3)$ of $3\times 3$ rotation matrices with matrix multiplication and matrix inverse.
Note that the group action $\phi$ is right handed
\[
\phi_{Q_1} \circ \phi_{Q_2} (R) :=  Q_1^\top Q_2^\top  \eta = ( Q_2 Q_1 )^\top  \eta  = \phi_{Q_2 Q_1} (R).
\]

For the second order kinematics \eqref{eq:lin_kin} we will consider two separate symmetry groups in the present paper.
The first symmetry group considered is $\grpG = \grpR^3 \times \grpR^3$ under addition.
That is $(\alpha_1,\beta_1) (\alpha_2, \beta_2) = (\alpha_1 + \alpha_2, \beta_1 + \beta_2)$ and $(\alpha,\beta)^{-1} = (-\alpha,-\beta)$.
This group acts on $\calM = \R^m \times \R^m$ via addition
\begin{align}
\phi_{(\alpha,\beta)} (p, v) & = (p + \alpha, v + \beta).
\label{eq:linhi}
\end{align}
This group action captures the Galilean transformations that correspond to expressing linear kinematics with respect to frame of reference that is translated and also moving with an arbitrary linear velocity.
The group action is commutative $\phi_{(\alpha_1,\beta_1)} \phi_{(\alpha_2,\beta_2)} = \phi_{(\alpha_2,\beta_2)} \phi_{(\alpha_1,\beta_1)}$ and the concept of handedness does not apply in this case.
Later in the paper (Section~\ref{sec:outputs}) we will introduce a second symmetry group to show that the symmetry is not unique and demonstrate the importance of choosing symmetries that respect the sensors as well as the kinematics.

The concept of equivariance of a control system has been around since at least the eighties \cite{1985_Grizzle_StructureNLSym,1981_VdSchaft_SymmetryHamilton,1985_Nijmeijer_PartialSyms}.
The modern formulation is present in the work of Aghannon \etal \cite{Aghannan2003} and is core to the work on moving frame observers from that period \cite{Bonnabel2006_acc,2008_martin_InvariantObserverEarthVelocityAided,bonnabel2009}.
This formulation is used in the authors previous work observer for systems on homogeneous spaces \cite{RM_2013_Mahony_nolcos}.
The present development introduces a new perspective on the same structure: we use the state-space symmetry to define a group action on the set of vector fields \cite{2020_Mahony_EquivariantSystemsTheory} and in turn use this to develop a unified geometric interpretation of equivariance and invariance.

A symmetry $\phi_X$, a diffeomorphism of the state-space $\calM$, naturally induces a linear mapping on the set of vector fields $\gothX(\calM)$ on $\calM$ via the push forward operation.
Let $\tD \phi_X : \tT_\xi \calM \to \tT_{\phi_X(\xi)} \calM$ be the differential of a symmetry $\phi_X$.
The \emph{push forward} map $\Phi_X : \gothX(\calM) \to \gothX(\calM)$ defined by
\begin{align*}
\Phi_X (f)  & := \tD \phi_X \circ f \circ \phi_X^{-1}
\label{eq:td_star}
\end{align*}
is a well defined linear map that has the structure of a group action
$\Phi : \grpG \times \gothX(\calM) \to \gothX(\calM)$, $\Phi(X,f) := \Phi_X (f)$  \cite{2020_Mahony_EquivariantSystemsTheory}.
Since $\Phi_X$ acts on the vector fields it can be applied to the system function $f$ and provides a natural structure to define invariance and equivariance.

A vector field $f \in \gothX(\calM)$ is \emph{invariant} if
\begin{align*}
  \Phi_X (f) = f
\end{align*}
for all $X \in \grpG$.
That is, a vector field is invariant if $\tD \phi_X f(\xi) = f(\phi_X(\xi))$.
Similarly, a system $f : \vecL \to \gothX(\calM)$ is \emph{invariant} (with respect to $\Phi$) if it satisfies
\begin{align}
\Phi_X (f_{u}) = f_{u}
\end{align}
for all $X \in \grpG$ and $u \in \vecL$.
Explicitly one has $\tD \phi_X f(\xi,u) = f(\phi_X(\xi),u)$ for all $\xi \in \calM$, $u \in \vecL$ and every $X \in \grpG$.
Note that an invariant system is made up of invariant vector fields.
In particular, for a constant input $u \in \vecL$ the associated vector field $f_{u}$ is independently invariant.

Consider the linear kinematics \eqref{eq:lin_kin} and the Galilean group action \eqref{eq:linhi}.
The input vector fields for this system are $f^{\text{ctl}}_{a}(\xi) = (0, {a})$.
One has
\[
\Phi_{(\alpha,\beta)} (f^{\text{ctl}}_{a}(p, v))
= \tD \phi_{(\alpha,\beta)} f^{\text{ctl}}_{a}(p-\alpha,v-\beta)
= \tD \phi_{(\alpha,\beta)} (0, a)
= (0, a)
= f^{\text{ctl}}_{a}(p, v).
\]
As one would expect, the linear input function $f^{\text{ctl}}_{a}$ is invariant to the Galilean symmetry action.
For the drift field $f_{0}(p, v) = (v, 0)$ one has
\[
\Phi_{(\alpha,\beta)} (f_{0}(p, v)) =
\tD \phi_{(\alpha,\beta)} f_{0}(p-\alpha, v-\beta)
= \tD \phi_{(\alpha,\beta)} (v-\beta,0)
= (v-\beta,0)
\not= f_{0}(p, v),
\]
and this is not an invariant vector field.
The constant velocity offset for the reference frame associated with the Galilean symmetry breaks symmetry in the position kinematics (the $\dot{p}$ kinematics).
Note that the velocity kinematics (the $\dot{v}$ kinematics) are invariant as would be expected by Newton's laws of motion.

\begin{figure}
  \centering
  \includegraphics[scale=0.56]{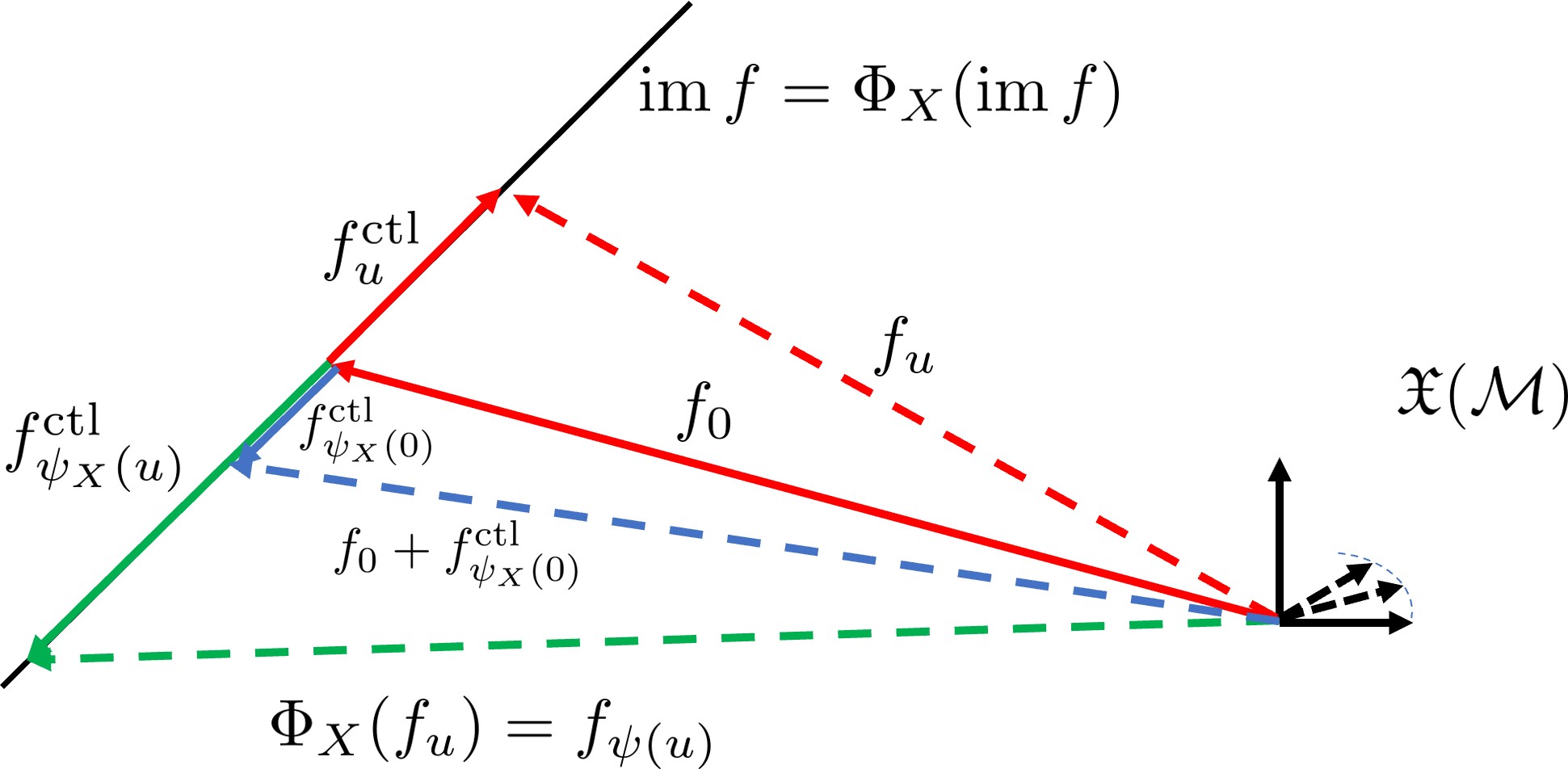}
  \caption{The affine subspace $\image f = \{f_{u} \in \gothX(\calM) \;|\; u \in \vecL \}$.  For an equivariant system then $\image f$ is fixed by the induced group action $\Phi : \grpG \times \gothX(\calM) \to \gothX(\calM)$. }
  \label{fig:image_f}
\end{figure}

A system is \emph{equivariant} if $f : \vecL \to \gothX(\calM)$ satisfies
\begin{align}
\Phi_X (f_{u}) = f_{\psi_X(u)}
\label{eq:equivariance_f}
\end{align}
for an \emph{input group action} $\psi :\grpG \times \vecL \to \vecL$.
Written in the classical form one has $\tD \phi_X f(\xi,u) = f(\phi_X(\xi),\psi_X(u))$ \cite{1985_Grizzle_StructureNLSym,2008_Bonnabel_TAC,bonnabel2009,RM_2013_Mahony_nolcos,2020_Mahony_EquivariantSystemsTheory}.
That is, the action $\Phi_X$ takes the vector field $f_{u}$ to another vector field $f_{\psi_X(u)}$ that \emph{also lies in the system image}.
The situation is visualised in Figure \ref{fig:image_f}.
The image of $f$ is a finite dimensional affine subspace of the infinite dimensional linear space $\gothX(\calM)$.
The action $\Phi$ is well defined on the infinite dimensional space $\gothX(\calM)$; indeed, each $\Phi_X : \gothX(\calM) \to \gothX(\calM)$ is a linear automorphism of $\gothX(\calM)$.
Equivariance captures the property that $\Phi_X$ maps $\image f$ \emph{into} $\image f$.
In prior work, the input action $\psi : \grpG \times \vecL \to \vecL$ was regarded as a separate structure \cite{1985_Grizzle_StructureNLSym,2008_Bonnabel_TAC,bonnabel2009,RM_2013_Mahony_nolcos}.
However, with this perspective it is clear that the substantive condition is that the $\Phi$ on $\gothX(\calM)$ preserves the affine subspace $\image f$ and that, once this is true, the action $\psi$ is fully defined by $\Phi$ through \eqref{eq:equivariance_f}.
The property that $\psi :\grpG \times \vecL \to \vecL$ is a group action follows from the group action property of $\Phi$, as long as it is closed on $\image f$.
Due to the affine structure of the system $f$, the input action $\psi_X$ is also affine; that is, $\psi_X(0) \not= 0$ in general.
Indeed, $\psi_X(0) = 0$ if and only if the drift vector field $f_0(\xi)$ is invariant $\Phi_X(f_0) = f_0$.

Consider the attitude kinematics \eqref{eq:direction_kinematics} with group action \eqref{eq:atthi}.
One has
\[
\Phi_{Q} (\Omega \times \eta) =
\tD \phi_{Q} (-\Omega \times \phi_{Q^{-1}}(\eta)) = Q^\top (-\Omega \times Q \eta) = -(Q^\top \Omega) \times \eta.
\]
Direction attitude kinematics are equivariant and the associated group action on the input $\Omega$ is given by $\psi : \SO(3) \times \R^3 \to \R^3$
\begin{align*}
\psi(Q,\Omega) := Q^\top \Omega.
\end{align*}
Note that $\psi$ is a right-handed group action $\psi_{Q_1} ( \psi_{Q_2} (\Omega)) = Q_1^{-1} Q_2^{-1}\Omega = (Q_2 Q_1)^{-1} \Omega = \psi_{Q_2 Q_1} (\Omega)$.

Returning to second order linear kinematics \eqref{eq:lin_kin}.
Consider an extension of the kinematics
\begin{subequations}
  \label{eq:lin_kin_ext}
  \begin{align}
    \dot{p} & = v + w \\
    \dot{v} & = a
  \end{align}
\end{subequations}
to include a new virtual input $w \in \R^3$ that can be chosen arbitrarily \cite{2020_Mahony_EquivariantSystemsTheory,vanGoor2020_EqF.cdc,vanGoor2020_EqF.tac}.
The original linear kinematics are recovered by setting $w = 0$, so there is no loss of generality in considering this more general system.
For the new system the input space is $\vecL = \R^3 \times \R^3$ with elements $(w,a)$.
The input function for the extended system can be written  $f^{\text{ctl}}_{(w,a)}(p,v) = (w, a)$ and specialises to the input function for the old system by setting the input to $(0,a)$.

As discussed earlier, the drift vector field $f_{0} = (v,0)$ was not invariant to transformation by $\Phi$ for the second order kinematics.
However, by including the virtual input the extended system becomes equivariant with input action $\psi : \grpG \times \vecL \to \vecL$
\begin{align*}
  \psi((\alpha, \beta), (w, a)) := (w - \beta, a).
\end{align*}
To see this compute
\begin{align*}
  \Phi_{(\alpha, \beta)} (f_{(w, a)} (p, v))
  &= \tD \phi_{(\alpha, \beta)} f_{(w, a)} \circ \phi_{(\alpha, \beta)}^{-1}(p, v), \\
  &= \tD \phi_{(\alpha, \beta)} f_{(w, a)} (p - \alpha, v - \beta), \\
  &= \tD \phi_{(\alpha, \beta)} [v - \beta + w, a], \\
  &= (v - \beta + w, a), \\
  &= f_{(w -\beta, a)}(p, v), \\
  &= f_{\psi((\alpha, \beta), (w, a))}(p, v).
\end{align*}
The velocity offset introduced by the Galilean symmetry that broke invariance is captured in the equivariance of the input through the virtual input $w$.
A way to interpret this structure physically is as modelling a constant velocity offset of the reference frame as an exogenous input rather than as a separate parameter in the modelling of the system.

Recalling Figure~\ref{fig:image_f}, the structure of adding virtual inputs is easily visualised.
Consider firstly the input vector fields $f^{\text{ctl}}_u(\xi)$.
Let $L^{\text{ctl}} = \spn \{f^{\text{ctl}}_u \;|\; u \in \vecL \} \subset \gothX(\calM)$ and note that $\image f = f_0 + L^{\text{ctl}}$.
Define a linear subspace of $\gothX(\calM)$ by
\[
\Phi_X L^{\text{ctl}} = \spn \{\Phi_X (f^{\text{ctl}}_u) \;|\; X \in \grpG, u \in \vecL\}
\]
Similarly, define the linear subspace $\Phi_X L^{\text{drift}} = \spn\{\Phi_X(f_0) - f_0\;|\; X \in \grpG\}$ in $\gothX(\calM)$.
Define
\[
L^{\text{ctl}}_{\text{ext}} = \Phi_X L^{\text{ctl}} \cup  \Phi_X L^{\text{drift}}.
\]
Geometrically, the affine set $\image f$ in Figure~\ref{fig:image_f} is extended to a new affine set $f_0 + L^{\text{ctl}}_{\text{ext}}$ with the same offset $f_0$ and a larger linear span.
By construction this new affine set contains all possible vector fields that can be obtained by acting on the old system by $\Phi_X$, including the original system since $\Phi_{\Id} = \id$.
Moreover, since $\Phi_X \Phi_Y = \Phi_{YX}$ is a group action, then this new affine set is closed under action by $\Phi_X$.
Choose a basis $\{ f^1_i \} \in \gothX(\calM)$ for $L^{\text{ctl}}_{\text{ext}}$ that contains the input vector fields $\{f_i\}_{i = 1}^l$ of the original system.
Define the extended system $f^{\text{ext}}_w = f_0 + \sum_{j = 1}^\infty w^{j} f^1_{j}$ (for $w$ with finitely many non-zero elements arbitrary), noting that the old system $f_u = f_0 + \sum_{i = 1}^l u^i f_i$ is wholly contained in the image of the extended system.
The group action $\psi : \grpG \times L^{\text{ctl}}_{\text{ext}} \to L^{\text{ctl}}_{\text{ext}}$ is defined implicitly by the action $\Phi_X$ since $\image f^{\text{ext}}_w$ maps into $\image f^{\text{ext}}_w$.
Of course, an extended system constructed in this way may in general be infinite dimensional, however, in many cases of interest, such as the Galilean system considered as an example in this paper, the extension is finite.
This construction provides a clear and highly practical methodology to apply equivariant observer design to a wide range of systems defined on homogeneous spaces including certain systems that were not equivariant as originally formulated.

For a transitive group action then for all $\mr{\xi}$ and $\xi \in \calM$ there is an $X \in \grpG$ such that $\phi_X(\mr{\xi}) = \xi$.
In particular, the map $\phi_{\mr{\xi}} :  \grpG \to \calM$,
\begin{align*}
\phi_{\mr{\xi}}(X) := \phi(X,\mr{\xi})
\end{align*}
is a submersion.
Fixing a reference point $\mr{\xi} \in \calM$, the group $\grpG$ parametrizes $\calM$, that is, one can represent any element $\xi$ of the manifold $\calM$ by the image $\xi = \phi(X,\mr{\xi})$ of an element $X \in \grpG$.
Although straightforward this point deserves more attention as it one of the key concepts in equivariant observer design.
The parametrization provided by the group allows the observation problem to be formulated at the group level rather than directly on the state-manifold.
Rather than trying to find an estimate $\hat{\xi} \in \calM$ for the true state $\xi \in \calM$, one can search for an estimate $\hat{X} \in \grpG$ such that $\hat{\xi} = \phi(\hat{X}, \mr{\xi})$ is an estimate of the state.
This simple reparametrization is more powerful than it may appear at first glance.
Abstractly, we now pose the observer state on the symmetry group $\hat{X} \in \grpG$ and the state estimate becomes an output generated by the action of the observer state on an arbitrary (but fixed) \emph{origin} point $\mr{\xi} \in \calM$.
The observer state, as an element of the symmetry group, has much more structure than just the state estimate; indeed, it corresponds to a symmetry $\phi_{\hat{X}}$ while $\hat{\xi} \in \calM$ is just a point.
In particular, $\phi_{\hat{X}}$ is a diffeomorphism on $\calM$ and can be applied to any point in the space, not just the origin $\mr{\xi}$.
This structure is fundamental in the construction of a globally defined intrinsic error (cf.~Section \ref{sec:Observer}) that is the foundation of the observer design methodology.

\begin{figure}
  \centering
  \includegraphics[scale=0.416]{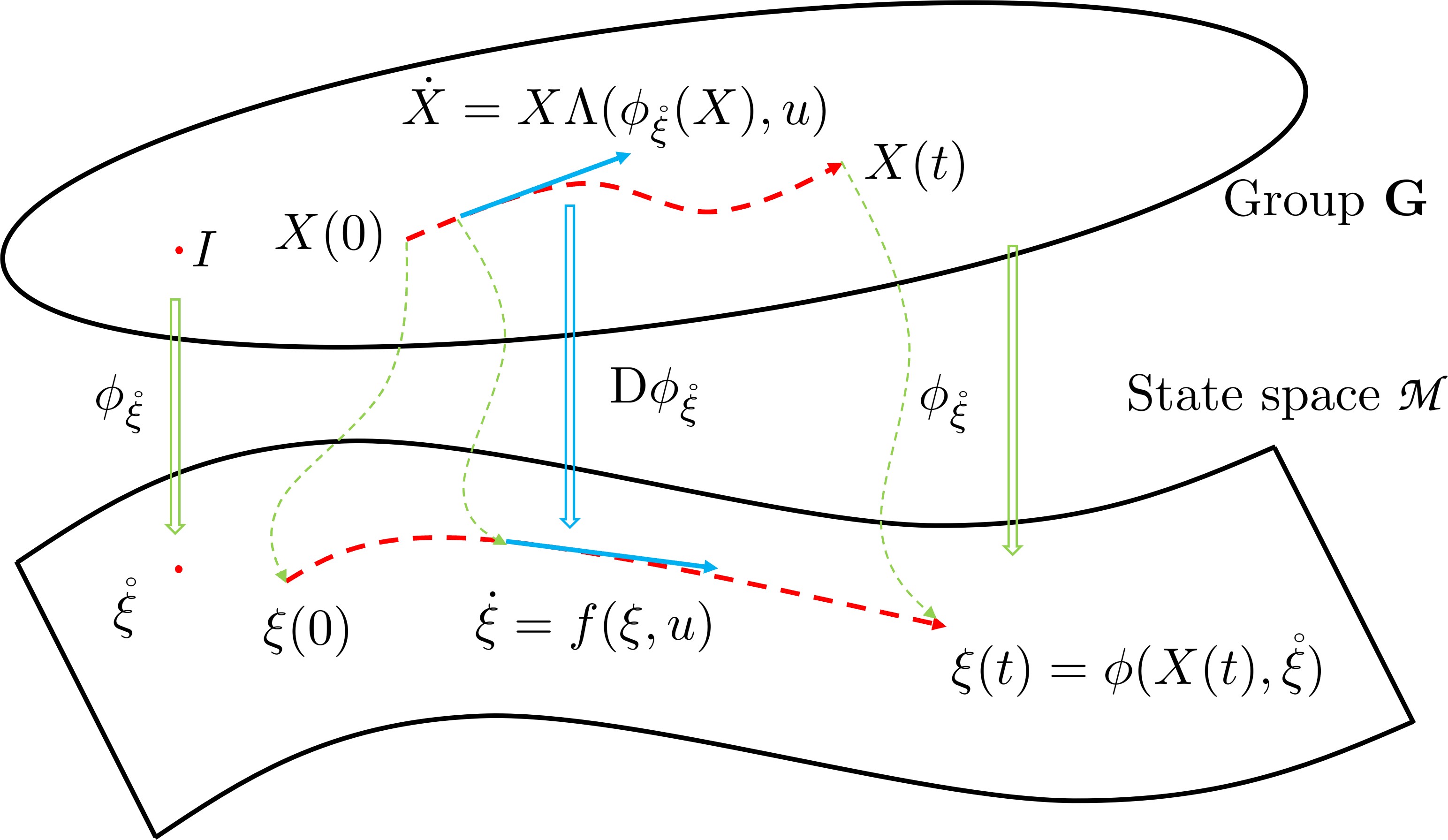}
  \caption{The lifted system on $\grpG$ evolves such that its projection $\phi_{\mr{\xi}}(X(t) \in \calM$ is a solution of the system \eqref{eq:system}.}
  \label{fig:LiftedSystem}
\end{figure}

Choosing the observer state on the symmetry group allows for a globally defined intrinsic error (as we will show in Section~\ref{sec:Observer}), however, it poses a separate challenge in observer design; that of choosing internal model dynamics for the observer.
Internal model dynamics are the unforced dynamics of the observer that enable it to continue to track the system state when the measurement has converged.
For a classical observer, where the observer state space is the same as the system-state space, the internal dynamics are chosen to be a copy of the system dynamics.
If the initial observer state and system state are equal, and the observer and system model are fed with the same inputs, the two trajectories will be equal, at least up to perturbations due to disturbances and modelling error.
In contrast, the symmetry group may have a higher dimension than the state-manifold and although there is a natural projection $\phi_{\mr{\xi}} : \grpG \to \calM$ the inverse of this map is not intrinsic and defining a good internal model on the group is not straightforward.
The stabiliser $\stab_\phi(\mr{\xi})$  of a point $\xi \in \calM$ is the set $\{S \in \grpG \;|\; \phi_S(\mr{\xi}) = \mr{\xi}\}$ that captures the additional degrees of freedom in $\grpG$ compared to $\calM$.
Defining an internal model on $\grpG$ boils down to choosing a set of compatible dynamics in the stabiliser to complement dynamics on $\grpG$ induced by the system, such that resulting trajectories on $\grpG$ project down to the system trajectories on $\calM$ as shown in Figure~\ref{fig:LiftedSystem}.
Note that since there is no intrinsic global factorisation of the symmetry group of a homogeneous space into stabiliser and manifold directions there is no simple way of separating the choice of stabiliser dynamics from the induced dynamics and the internal model must be chosen to integrate both requirements into a single set of system dynamics defined on the symmetry group.

A key advance in equivariant systems theory \cite{2020_Mahony_EquivariantSystemsTheory} was the understanding that a lifted system (of an equivariant system on $\calM$) can always be written
\begin{align}
\dot{X} = X \Lambda(\phi_{\mr{\xi}}(X), u), \quad X(0) \in \grpG
\label{eq:lifted_system}
\end{align}
where $\Lambda : \calM \times \vecL \to \gothg$ is termed a \emph{lift function} \cite{2020_Mahony_EquivariantSystemsTheory}.
The lifted system formulation \eqref{eq:lifted_system} encodes internal model dynamics in the standard left invariant form $\dot{X} = X \Lambda$ where $\Lambda \in \gothg$ is an element of the Lie-algebra of the Lie-group and $X \Lambda = \tD L_X \Lambda$ is the left translated tangent vector element of the tangent space of $\grpG$ at $X$ \cite{Bullo2005_GeometricControlBook,Jurdjevic1997_GeometricControlTheory}
where $L_X Y = XY$ is the left multiplication on the group.
The system lift formulation is more powerful than considering a general $\dot{X} = X U(X,u)$ structure for $U : \grpG \times \vecL \to \gothg$,
since $\Lambda$ as a map defined on $\calM \times \vecL$ ensures that the resulting lifted dynamics are in some sense independent of the stabiliser on $\grpG$.
In particular, if $S \in \stab_\phi(\mr{\xi})$ then $\Lambda(\phi_{\mr{\xi}}(S),u) = \Lambda(\mr{\xi},u)$.
This ensures that the internal model dynamics of the observer are invariant to left translations $X \mapsto S X$ by the stabiliser
\[
\ddt SX = S \dot{X}
= SX \Lambda (\phi_{\mr{\xi}}(SX) , u)
= SX \Lambda (\phi_X (\phi_S (\mr{\xi})),u)
= SX \Lambda (\phi_{\mr{\xi}}(X),u).
\]

The system lift (eq.~\ref{eq:lifted_system}) is chosen to satisfy the following two conditions: \\
\noindent{\textbf{Pre-image:}}
A necessary and sufficient condition to ensure that solutions of the lifted system project to solutions of the actual system is that
\begin{align}
\tD \phi_\xi \Lambda(\xi, u) = f(\xi,u)
\label{eq:project_Lambda}
\end{align}
where $\tD \phi_\xi : \gothg \to \tT_\xi \calM$ is the differential of the group action in its group variable, for a given $\xi \in \calM$, evaluated at the identity in $\grpG$ \cite{RM_2013_Mahony_nolcos,2020_Mahony_EquivariantSystemsTheory}.
From this it is straightforward to prove that as long as $\phi_{\mr{\xi}}(\hat{X}(0)) = \xi(0)$ then trajectories of \eqref{eq:lifted_system} project to trajectories of \eqref{eq:system} (Figure~\ref{fig:LiftedSystem}).
Since the map $\phi_{\mr{\xi}} : \grpG \to \calM$ is a submersion the differential $\tD \phi_{\mr{\xi}} : \gothg \to \tT_\xi \calM$ is surjective, and it is always possible to define a pre-image $\Lambda(\xi,u)$ of $f(\xi,u)$ in $\gothg$ \cite{RM_2013_Mahony_nolcos}.
For a free group action, where the stabiliser is trivial and hence $\phi_{\xi} : \grpG \to \calM$ is invertible,  then $\Lambda(\xi,u) = \tD \phi_{\xi}^{-1} f(\xi,u)$ and $\Lambda$ is fully defined by \eqref{eq:project_Lambda}.
Such lifts naturally lead to equivariant system dynamics on the symmetry group.
Indeed, for a free action the state-manifold is isomorphic to the symmetry group and the design problem specialises to the case of equivariant observer design on a Lie-group.
In general, when the symmetry is not a free action, the pre-image condition \eqref{eq:project_Lambda} is not sufficient to ensure that the lifted system has symmetry properties on the group and without an additional condition leads to issues in the observer analysis.

\noindent{\textbf{Equivariance:}}
The second condition required is that the lift respects the symmetry.
This condition is encoded as \cite{2020_Mahony_EquivariantSystemsTheory}
\begin{align}
\Ad_{X^{-1}} \Lambda(\xi,u) = \Lambda(\phi_X(\xi), \psi_X(u))
\label{eq:equivariant_lift_Lambda}
\end{align}
where $\Ad_X : \gothg \to \gothg$, $U \mapsto X U X^{-1}$, is the adjoint automorphism of the Lie-algebra \cite{Bullo2005_GeometricControlBook,Jurdjevic1997_GeometricControlTheory}.
The role of the adjoint here is as the symmetry transformation of the Lie-algebra associated with change of base point $\xi \in \calM$.
In particular, for any homogeneous space the following diagram is known to commute \cite{Bullo2005_GeometricControlBook,Jurdjevic1997_GeometricControlTheory}
\[
\xymatrix{
	\gothg \ar@{->}[r]^{\Ad_{X^{-1}}} \ar@{->}[d]_{\tD \phi_{\mr{\xi}}} &
	\gothg \ar@{->}[d]^{\tD \phi_{\phi_X (\mr{\xi})}} \\
	\tT_{\mr{\xi}} \calM \ar@{->}[r]^{\tD \phi_{X}} &
	\tT_{\phi_X (\mr{\xi})} \calM
	}
\]
Recalling \eqref{eq:project_Lambda}, it is natural to lift the equivariance condition $\tD \phi_X f(\xi, u) = f(\phi_X(\xi), \psi_X(u))$ for the system on $\calM$ to \eqref{eq:equivariant_lift_Lambda} on the symmetry group.
For a free group action, $\Lambda(\xi,u) = \tD \phi_{\xi}^{-1} f(\xi,u)$ (from \eqref{eq:project_Lambda}) and equivariance of the lift can be proved by exploiting the equivariance of the system model.
However, for groups with non-trivial stabiliser, the condition \eqref{eq:equivariant_lift_Lambda} extends this relationship to the stabiliser.
That is, for any  $S \in \stab_\phi(\xi)$ then \eqref{eq:equivariant_lift_Lambda} becomes
$\Ad_{S^{-1}} \Lambda(\xi,u) = \Lambda(\xi, \psi_S(u))$ since $\phi_S(\xi) = \xi$ by construction.
It is only in the unusual case where the stabiliser is a normal subgroup of the symmetry group that this constraint can be ignored without compromising the error dynamics defined in Section~\ref{sec:Observer}.
Existence of an equivariant lift for any equivariant system is a non-trivial question that was answered in the affirmative in a recent work \cite{2020_Mahony_EquivariantSystemsTheory}.
There is a construction given in \cite{2020_Mahony_EquivariantSystemsTheory} for building a system lift for an arbitrary system based on an iterative process that progressively works through a basis for the input space $\vecL$.
The tricky part is that for each new direction in $\vecL$, the behaviour of the lift on the stabiliser must be constructed to satisfy the equivariance condition and this influences the next choice of a direction in $\vecL$.
A full discussion is beyond the scope of the present paper beyond noting that, in practice, the most effective way of determining a lift known to the authors is to use \eqref{eq:project_Lambda} and then make educated guesses that can be verified using \eqref{eq:equivariant_lift_Lambda}.

The lifted system in the case of the direction kinematics is simply
\[
\dot{R} = R \Omega^\times
\]
where $\Omega^\times \in \so(3)$ is in the Lie-algebra of $\SO(3)$, the set of skew-symmetric matrices.
The system lift $\Lambda(\eta,\Omega) = \Omega^\times$ maps the input directly to the Lie-algebra with no state dependence.
The simplicity of this correspondence is both a boon and a curse, as it makes observer design simple for direction kinematics but also hides much of the structure of the general case.

For second order linear kinematics, consider a trajectory $(p(t), v(t)) = \phi_{(\alpha,\beta)} (\mr{p},\mr{v})$ associated with a lifted system trajectory $(\alpha(t),\beta(t))$ and constant origin $(\mr{p}, \mr{v}) \in \R^3 \times \R^3$.
One has
\[
\ddt \phi_{(\alpha,\beta)} (\mr{p},\mr{v}) = (\dot{p},\dot{v}) = (v + w, a)
\]
Thus the system lift is $\Lambda((p,v),(w,a)) := (v + w, a)$.
Note that the system lift function depends on the system state.
To compute the lifted system in terms of $(\alpha, \beta)$ one computes $\ddt (\alpha, \beta) = \tD L_{(\alpha,\beta)} \Lambda(\phi_{(\alpha,\beta)}(\mr{p}, \mr{v}),(w,a))$ and sets $w = 0$ (corresponding to the actual input) to yield
\begin{align*}
\dot{\alpha} & = \mr{v} + \beta \\
\dot{\beta} & = a
\end{align*}
The offset $\mr{v}$ in the first equation is important since the symmetry structure of $\beta$ encodes the offset between the true velocity and reference $\beta = v - \mr{v}$ and not directly the system velocity.

\section{Equivariant Error and the Observer Architecture}\label{sec:Observer}

The classical observer error $\tilde{\xi} = \xi - \hat{\xi} \in \R^m$ is defined for systems with Euclidean state space $\xi \in \R^m$ and where the observer is a copy of the system $\hat{\xi} \in \R^m$.
This error has been the foundation of observer design both for nonlinear constructive design methods, where $\tilde{\xi}$ is either the primary variable in a Lyapunov construction or is fundamental in the stability analysis, as well as for linearising design methods, where the error $\tilde{\xi}$ is linearised around $\tilde{\xi} = 0$.
For systems on manifolds, $\tilde{\xi}$ can only be constructed using an atlas of local or embedded coordinates centered around the time-varying state estimate $\hat{\xi}(t)$.
This approach to defining an observer error is neither intrinsic nor global.
In contrast, a key step in equivariant observer design is the construction of an equivariant error that is both intrinsic and globally defined.
Recent work \cite{2020_Mahony_EquivariantSystemsTheory} showed that this is only possible by exploiting the symmetry structure of the problem.
In particular, there is no error construction $e : \calM \times \calM \to \calM$, $e(\hat{\xi},\xi) \in \calM$ between observer and system states at the manifold level \cite[Theorem 5.7]{2020_Mahony_EquivariantSystemsTheory} that has the nice properties that we will demonstrate with the equivariant error defined below (Eq.~\eqref{eq:error_dynamics_equivariant}).
The equivariant error construction (Eq.~\eqref{eq:e}) overcomes this structural constraint by posing the observer state on the symmetry group.
That is, the equivariant error is a map $e: \grpG \times \calM \to \calM$, $e (\hat{X}, \xi) \in \calM$ between an element of the symmetry group $\hat{X} \in \grpG$ and the system state $\xi \in \calM$.
This construction motivates the choice to pose the observer state on the symmetry group and in turn motivates the work done in Section~\ref{sec:symmetry} to build a lifted system on $\grpG$ to act as  internal model for the observer dynamics.

For an observer state $\hat{X} \in \grpG$ and a system state $\xi \in \calM$ the \emph{equivariant error} is defined to be
\begin{align}
e := \phi_{\hat{X}^{-1}} (\xi) \in \calM.
\label{eq:e}
\end{align}
To understand this error recall that the state estimate is given by $\hat{\xi} = \phi_{\hat{X}} (\mr{\xi})$ for a fixed origin $\mr{\xi}$.
If $e = \mr{\xi}$ it follows that
\begin{align}
\hat{\xi} = \phi_{\hat{X}} (\mr{\xi}) = \phi_{\hat{X}} (e) =\phi_{\hat{X}} (\phi_{\hat{X}^{-1}} (\xi)) =\xi.
\label{eq:e2mrxi}
\end{align}
Thus, driving the error $e \to \mr{\xi}$ ensures the that state estimate $\hat{\xi} = \phi_{\hat{X}} (\mr{\xi})$ converges to the true state $\xi$.

It is instructive to consider the linear kinematics example.
The observer state $\hat{X} \in \grpR^3 \times \grpR^3$ is an element of the Galilean group.
Write $\hat{X} = (\hat{\alpha}, \hat{\beta})$, then
\[
e = \phi(\hat{X}^{-1},\xi) = \phi_{(-\hat{\alpha},-\hat{\beta})}(p,v) = (p - \hat{\alpha}, v  - \hat{\beta}).
\]
Choose $\mr{\xi} = (0, 0)$ to be the origin point in $\R^3 \times \R^3$ and define the state estimate to be $(\hat{p}, \hat{v}) = \phi_{(0,0)} (\hat{\alpha}, \hat{\beta}) = (\hat{\alpha}, \hat{\beta})$.
Substituting into the error yields $e = (p - \hat{p}, v  - \hat{v})$
the classical linear observer error.
Driving $e \to (0,0)$ is the same is driving $\tilde{p} = p - \hat{p}$ and $\tilde{v} = v - \hat{v}$ to zero.
The equivariant error construction passes the ``pub test'' of specialising to the classical linear error in the case where the state space $\calM = \R^m$ is linear.

The direction kinematics example, however, demonstrates the power of the equivariant error.
In this case $\hat{X} = \hat{Q} \in \SO(3)$ and there is no natural origin $\mr{\xi} = \mr{\eta}$.
Let us choose $\mr{\eta} = \eb_3$ in lieu of any better choice.
The equivariant error is
\[
e := \phi_{\hat{Q}^\top} (\eta) = \hat{Q} \eta \in \Sph^2.
\]
If $e = \eb_3 = \hat{Q} \eta$ then $\eta = \hat{Q}^\top \eb_3 = \hat{\eta}$ where the second equality is the definition of $\hat{\eta} = \phi_{\hat{Q}}(\eb_3)$.
In contrast, the difference in state elements $(\eta - \hat{\eta})$ is not an element of $\Sph^2$ and can only be expressed in a non-intrinsic way in either embedded or local coordinates.
More importantly, the equivariant error lives in a neighbourhood of a fixed origin $\mr{\eta} = \eb_3$ on the sphere while the coordinates for the difference $(\eta - \hat{\eta})$ must be chosen carefully and changed periodically as $\eta$ and $\hat{\eta}$ move.

For any $\hat{X}$, the equivariant error $e = \phi_{\hat{X}^{-1}}(\xi) \in \calM$ is well defined.
This opens the door to global analysis of observer convergence, something  that is rarely possible for standard observer architectures applied to nonlinear systems on manifolds.
Nonlinear observer designs can be designed based on Lyapunov functions $V : \calM \to \R_+$ written as a function of error coordinates $V(e)$ and centered at the origin $V(\mr{\xi}) = 0$.
Construction of the Lyapunov functions and design of the observer must be undertaken on a case-by-case basis, however, the resulting observers are inherently globally defined and their basins of attraction are only limited by the construction of the Lyapunov function and topological constraints of the manifold $\calM$
\cite{
RM_2011_Hua_cdc,
2010_vasconcelos_NonlinearPositionAttitude,
2011_madgwick_EfficientOrientationFilter,
2011_Hamel_cdc,
2012_Trumpf_TAC,
2012_grip_AttitudeEstimationUsing,
2012_batista_SensorBasedGloballyAsymptotically,
2014_izadi_RigidBodyAttitude,
2015_Hua,
2016_allibert_VelocityAidedAttitude,
2016_hua_StabilityAnalysisVelocityaided,
2017_berkane_HybridGlobalExponential,
2017_LeBras,
2015_Hua,
2018_Zlotnik_GradientSLAM,
MiaTayeb2019,
2019_hua_FeaturebasedRecursiveObserver,
2020_Hua_Automatica}.
For linearisation based observer design, the error dynamics can be linearised at $\mr{\xi}$ using a single set of local coordinates
\cite{
Bonnabel2006_acc,Martin2007,
2008_martin_InvariantObserverEarthVelocityAided,
2008_Bonnabel_TAC,
bonnabel2009,
2009_bonnabel_InvariantExtendedKalman,
RM_2012_Zamani_TAC,
bourmaud2013_discreteEKF,
2015_bourmaud_ContinuousDiscreteExtendedKalman,
2015_barrau_IntrinsicFilteringLie,
2016_Saccon_TRO,
2017_Barrau_tac,
2018_barrau_InvariantKalmanFiltering,
2019_Lavoie_InvariantHinf,
2020_Phogat_iEKF,
vanGoor2020_EqF.cdc,
vanGoor2020_EqF.tac}.
The improved regularity of the linearisation (no coordinate changes) leads to improved observer performances as is shown for the equivariant filter design presented in Section \ref{sec:design}.

\begin{figure}
  \centering
  \includegraphics[scale=0.416]{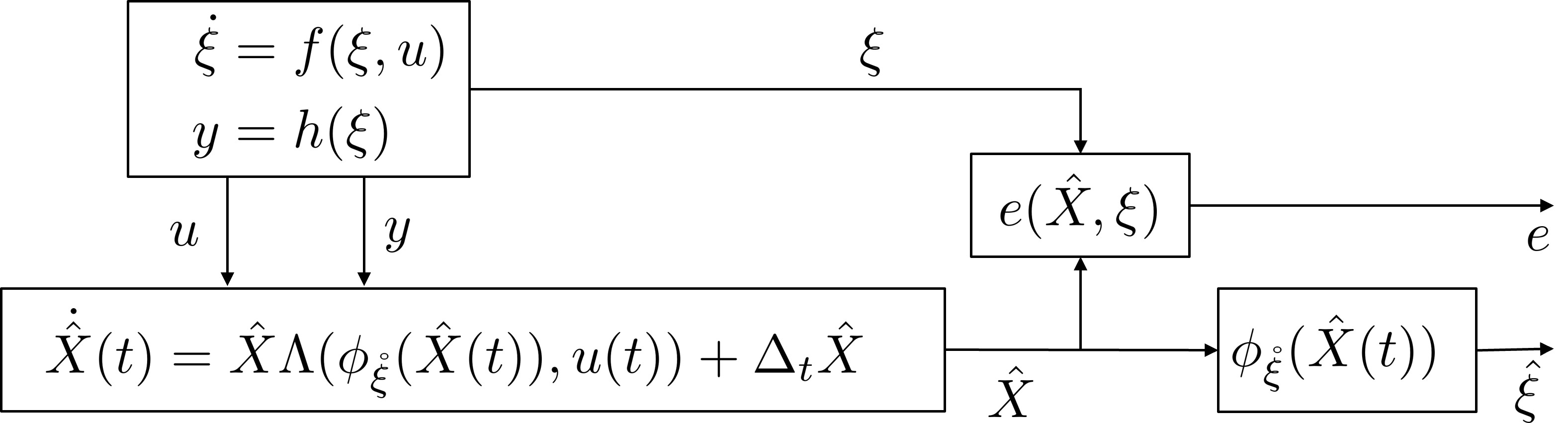}
  \caption{The Equivariant observer architecture.  Measurements from the system feed into the observer that is posed on the symmetry group.  The observer state and the system state feed into the equivariant error.  The observer is designed to drive $e \to \mr{\xi}$ and this ensures that the state estimate $\hat{\xi} \to \xi$ converges to the system state.
  }
  \label{fig:Observer}
\end{figure}

The equivariant error construction allows a principled observer architecture for equivariant systems as shown in Figure \ref{fig:Observer}.
The observer state lies in the symmetry group and uses the internal model provided by the lifted kinematics \eqref{eq:lifted_system}
\begin{align}
\ddt \hat{X} = \hat{X} \Lambda(\phi_{\mr{\xi}}(\hat{X}), u) -
\Delta_t \hat{X}.
\label{eq:observer}
\end{align}
Here the correction term $\Delta_t := \Delta_t(\hat{X},y) \in \gothg$ must be a function of the observed outputs $y$ and the known observer state $\hat{X}$ and remains to be defined.
The equivariant error $e$ is generated as a signal and the design goal is to drive $e \to \mr{\xi}$ asymptotically.
In the case that $e \to \mr{\xi}$ then the state estimate $\hat{\xi} = \phi_{\mr{\xi}} (\hat{X})$ will converge to the true state $\xi$ (see~\eqref{eq:e2mrxi}).
The observer state $\hat{X}$ lives in a higher dimensional space $\grpG$ and its convergence is not characterised directly.

The proposed architecture leads to a clear design principle: to design the correction term $\Delta_t$ such that the equivariant error dynamics $e \to \mr{\xi}$.
To do this one must study the dynamics of the equivariant error in detail.
It is a straightforward exercise to compute \cite{2021_mahony_HomogeneousSpaceGeometry}
\begin{align}
\ddt e(t) = \tD \phi_{e}\Ad_{\hat{X}} \left(\Lambda(\phi_{\hat{X}}(e),u) -  \Lambda(\phi_{\mr{\xi}} (\hat{X}), u) \right)
- \tD \phi_{e} \Delta_t.
\label{eq:error_dynamics_general}
\end{align}
Formally, the error dynamics are a system defined for  $e \in \calM$ with two exogenous inputs $u(t)$ and $\hat{X}(t)$ while the correction term can be considered separately as an arbitrary forcing term in the tangent space $\tT_e \calM$.
The input $u(t)$ dependence is structural in the problem, indeed, in many robotics problems a persistently exciting input signal is required to ensure observability of the system \cite{2012_Trumpf_TAC}.
However, the dependence of the error dynamics on $\hat{X}$ in such an embedded manner will lead to challenges in the analysis.
The fact that the evolution of $\hat{X}$ depends in turn on the choice of correction term $\Delta_t$ adds additional complexity.

This is where equivariance of the lifted system plays a critical role.
Recall that for an equivariant lift $\Ad_{\hat{X}}\Lambda(\xi,u) = \Lambda(\phi_{\hat{X}^{-1}}(\xi),\psi_{\hat{X}^{-1}}(u))$ (see~\eqref{eq:equivariant_lift_Lambda}).
Factoring the adjoint operator into the dynamics yields
\begin{align}
\ddt e(t)
& = \tD \phi_{e} \left(\Lambda(e,\mr{u}) -  \Lambda(\mr{\xi},\mr{u}) \right)
- \tD \phi_{e} \Delta_t
\label{eq:error_dynamics_equivariant}
\end{align}
where we introduce the notation
\begin{align}
\mr{u} :=\psi_{\hat{X}^{-1}}(u)
\label{eq:origin_velocity}
\end{align}
and term this the \emph{origin input}.
The best way to understand $\mr{u}$ is as a `change of basis' transformation that takes the input $u$ (a velocity measurement) and rewrites it in the correct `frame of reference' to apply to the error kinematics \eqref{eq:error_dynamics_equivariant}.
In robotic systems where sensors are mounted on the vehicle then the action $\psi_{\hat{X}^{-1}}$ is typically an explicit change of basis, usually a rotation, that transforms the measurement from the body-fixed-frame to the observers best estimate of the reference frame.
For example, in the direction kinematics example $\psi_{\hat{Q}^\top}(\Omega) = \hat{Q} \Omega$ can be interpreted as a change of basis for the input $\Omega$ measured in the body-fixed-frame to $\hat{Q} \Omega$ in the (observers best estimate of the) reference frame.

Whereas Eq.~\eqref{eq:error_dynamics_general} are dynamics in two exogenous inputs $\hat{X}$ and $u(t)$ as well as the correction term, the equivariant error dynamics Eq.~\eqref{eq:error_dynamics_equivariant} have a single exogenous input $\mr{u}(t)$ along with the correction term.
Analogous to the general case, the equivariant error dynamics often depend on a persistently exciting input $\mr{u}(t)$ to ensure observability, particularly for robotic systems.
This imposes some constraints on the relative relationship between the observer trajectory and the input, which, at least asymptotically, corresponds to a constraint on the relationship between the system trajectory and the raw input.
However, apart from the observability analysis, the fact that $\mr{u}(t)$ depends on the observer state makes no difference to the observer design.
The error dynamics \eqref{eq:error_dynamics_equivariant} are not autonomous (except in certain special cases \cite{RM_2013_Mahony_nolcos}), however, the simplicity and elegance of the equation provides a powerful tool in observer design.
For non-linear observer design then the global structure of $\Lambda(e,\mr{u})$ is of interest and there are many systems where this structure has tractable algebraic structure and it is possible to build globally defined non-linear observers
\cite{
RM_2011_Hua_cdc,
2010_vasconcelos_NonlinearPositionAttitude,
2011_madgwick_EfficientOrientationFilter,
2011_Hamel_cdc,
2012_Trumpf_TAC,
2012_grip_AttitudeEstimationUsing,
2012_batista_SensorBasedGloballyAsymptotically,
2014_izadi_RigidBodyAttitude,
2015_Hua,
2016_allibert_VelocityAidedAttitude,
2016_hua_StabilityAnalysisVelocityaided,
2017_berkane_HybridGlobalExponential,
2017_LeBras,
2015_Hua,
2018_Zlotnik_GradientSLAM,
MiaTayeb2019,
2019_hua_FeaturebasedRecursiveObserver,
2020_Hua_Automatica}.
For linearising observers then it is the local structure of the error dynamics that are of interest
\cite{
Bonnabel2006_acc,Martin2007,
2008_martin_InvariantObserverEarthVelocityAided,
2008_Bonnabel_TAC,
bonnabel2009,
2009_bonnabel_InvariantExtendedKalman,
RM_2012_Zamani_TAC,
bourmaud2013_discreteEKF,
2015_bourmaud_ContinuousDiscreteExtendedKalman,
2015_barrau_IntrinsicFilteringLie,
2016_Saccon_TRO,
2017_Barrau_tac,
2018_barrau_InvariantKalmanFiltering,
2019_Lavoie_InvariantHinf,
2020_Phogat_iEKF,
vanGoor2020_EqF.cdc,
vanGoor2020_EqF.tac}.
In this case, the linearisation of the unforced error dynamics in $e$ around $\mr{\xi}$ are straightforward to compute and provide a powerful model for observer design as we show in Section \ref{sec:design}.

\section{Equivariant outputs}\label{sec:outputs}

An observer problem is highly dependent on the measurements available.
An important design choice in equivariant observer design is to choose a symmetry, if possible, that is compatible with the output measurements.
Although this is not always possible, there are many recent examples where symmetries compatible with common sensor modalities are being discovered.
Interestingly, the corresponding system symmetries are often not the classical symmetries common in the physics literature.
This opens a whole line of research into what is the best symmetry to choose to analyse a systems and control problem and why.
We only touch on this question in the present paper by providing a motivating example.

An output $y = h(\xi)$ is equivariant if there exists a group action $\rho : \grpG \times \calN \to \calN$ such that
\begin{align}
\rho_{X}(y) = h(\phi_X(\xi)).
\label{eq:rho}
\end{align}
The existence of such a symmetry action is a structural constraint that involves the output and symmetry group.
For example, an $\R^3$ measurement of the position of a point $y = p = h(p,v)$ for the linear kinematics admits an output action
\begin{align*}
\rho_{(\alpha, \beta)} (y) = y + \alpha = p + \alpha = h(p + \alpha, v + \beta) = h (\phi_{(\alpha, \beta)} (p,v)).
\end{align*}
A similar output action exists for the direction kinematics output $y = \eta$.
In this case the output action is trivially that of the state space, since the full state has been measured
\begin{align*}
\rho_Q(y) = Q^\top \eta = \phi_Q(\eta).
\end{align*}
However, not all outputs are equivariant with respect to the commonly used symmetries.

For example, consider the case of the linear kinematics \eqref{eq:lin_kin} and imagine that there are separate bearing and range measurements of the position coordinate
\begin{align}
y_1 = \frac{p}{|p|} =: h_1(p,v) \in \Sph^2, \quad y_2 = |p| =: h_2(p,v) \in \R_{+},
\label{eq:y1y2}
\end{align}
a very common measurement modality in robotics.
Although these measurements could be (and commonly are) used to reconstruct the measurement $y = y_1 y_2 = p$, such an algebraic manipulation of the raw measurements distorts the noise properties of the model and impacts the performance of a filter or observer.
It is far preferable to consider a filter design that explicitly uses the raw measurements in the filter.
These raw measurements, however, do not admit a group action that is equivariant with respect to the Galilean symmetry.
It is important to note that this failure of equivariance is a structural constraint on \emph{both} the outputs \emph{and} the symmetry.
The outputs are part of the physical design of the system and cannot be changed without changing the sensor suite.
The symmetry, however, is a mathematical construct and it may be possible to deliberately choose a symmetry group for which the output does admit an equivariant group action.

To make this point clear, since it is one of the key differentiators of equivariant observer design, let us develop a symmetry that is compatible with bearing and range measurements for linear second order kinematics.
Consider the Lie-group $(\SO(3) \times \MR(1)) \ltimes \grpR^3$ with group multiplication
\begin{align}
(R_2,r_2,\beta_2)(R_1,r_1,\beta_1) = (R_2R_1,r_2r_1,\beta_2 + r_2 R_2 \beta_1).
\label{eq:polar_group_mult}
\end{align}
In addition to the special orthogonal group $\SO(3)$ of rotation matrices, we use the `multiplicative real' group $\MR(1)$ of all scalar positive reals under multiplication with identity 1, and the `additive real group' $\grpR^3$ of real 3-tuples under addition with identity $(0, 0, 0)$.
The notation `$\ltimes$' denotes a semi-direct product of $(SO(3) \times \MR(1))$ on $\grpR^3$ defined by the group multiplication \eqref{eq:polar_group_mult}.
We write the indices in `reverse' order to aid in verifying the action property \eqref{eq:polar_action} below.
The identity element of $(\SO(3) \times \MR(1))$ is $(I_3,1,0)$ and the inverse element is
\[
(R,r,\beta)^{-1} = \left(R^\top, \frac{1}{r}, - \frac{1}{r}R^\top \beta \right).
\]
Now consider a state symmetry $\phi' : ((\SO(3) \times \MR(1)) \ltimes \grpR^3) \times (\R^3 \times \R^3) \to (\R^3 \times \R^3)$,
\begin{align*}
\phi'((R,r,\beta),(p,v)) :=  \left(\frac{1}{r} R^\top p, \frac{1}{r} R^\top ( v - \beta)\right).
\end{align*}
It is straightforward to verify
\begin{align}
\phi'((R_1,r_1,\beta_1), \phi'((R_2,r_2,\beta_2),(p,v)))
& = \left(\frac{1}{r_2r_1} (R_2 R_1)^\top, \frac{1}{r_2r_1}(R_2 R_1)^\top ( v - (\beta_2 + r_2 R_2 \beta_1)) \right) \notag \\
& = \phi'((R_2,r_2,\beta_2)(R_1,r_1,\beta_1),(p,v))
\label{eq:polar_action}
\end{align}
and see that $\phi'$ is a group action on $\R^3 \times \R^3$.
This action is most certainly not a Galilean action.
It is, however, compatible with the bearing-range output sensor modality.
Consider output actions
\begin{gather*}
\rho^1_{(R,r,\beta)}(y_1) := R^\top y_1 = R^\top \frac{p}{|p|} =   \frac{\frac{1}{r} R^\top p}{\left|\frac{1}{r} R^\top p\right|} = h_1(\phi'_{(R,r,\beta)}(p,v)) \\
\rho^2_{(R,r,\beta)}(y_2) := \frac{1}{r} y_2 = \frac{1}{r} |p| = \left|\frac{1}{r} R^\top p\right| = h_2(\phi'_{(R,r,\beta)}(p,v)) \\
\end{gather*}

It remains to show that the proposed symmetry is compatible with the second order state kinematics.
Consider an input action $\psi'  :  \grpG \times \vecL \to \vecL$ defined by
\begin{align*}
\psi'((R,r,\beta),(w,a)) := \left(\frac{1}{r} R^\top (w + \beta), \frac{1}{r} R^\top a \right).
\end{align*}
Then the system is equivariant, since
\begin{align*}
  \Phi'_{(R, r, \beta)}(f_{(w,a)})(p,v)
  &= \tD \phi'_{(R, r, \beta)}f_{(w,a)}({\phi'}_{(R, r, \beta)}^{-1}(p,v)), \\
  &= \tD \phi'_{(R, r, \beta)}f_{(w,a)}\left(r R p, r R  v + \beta \right), \\
  &= \tD \phi'_{(R, r, \beta)}\left(w + r R  v + \beta, a \right), \\
  &= \left(\frac{1}{r} R^\top (w + r R  v + \beta), \frac{1}{r} R^\top a \right), \\
  &= \left(v + \frac{1}{r} R^\top (w + \beta), \frac{1}{r} R^\top a \right), \\
  &= f_{\psi'((R, r, \beta), (w, a))}(p, v).
\end{align*}

The advantage of an equivariant output is that it is now possible to define an equivariant innovation.
Recall the classical definition of an innovation $\tilde{y} = y - \hat{y}$
for $y, \hat{y} \in \R^m$.
The innovation $\tilde{y}$ captures the new information in the measurement $y$ compared to the existing knowledge $\hat{y} = h(\hat{\xi})$ of the state.
The equivariant innovation is defined to be
\begin{align}
d = \rho_{\hat{X}^{-1}} (y).
\label{eq:d}
\end{align}
In the same way that the classical innovation for a linear system satisfies $\tilde{y} = y - \hat{y} = C\xi - C \hat{\xi} = C\tilde{\xi}$ one has
\begin{align}
d = \rho_{\hat{X}^{-1}} (y) = \rho_{\hat{X}^{-1}} (h(\xi)) = h(\phi_{\hat{X}^{-1}} (\xi)) = h(e).
\label{eq:d_of_e}
\end{align}
In this sense the equivariant innovation captures the new information in the measurement $y$ with respect to the existing information on the state encoded in $\hat{X}$.
If the state estimate $\hat{X}$ is correct then $e = \mr{\xi}$ and $d = h(\mr{\xi}) = \mr{y}$ contains no new information on the state $\hat{X}$.
However, whereas an innovation $y - \hat{y}$ requires local or embedded coordinates and is neither globally defined nor intrinsic, the equivariant innovation provides a globally well defined intrinsic measure of new information.
This is an extremely important property for nonlinear observer design where global definition of the correction term is often one of the key design criteria.
It is also important in linearising observer design and leads to improved linearisation error in the equivariant filter \cite{vanGoor2020_EqF.tac}.

\section{Observer Design}\label{sec:design}

In this section we present a careful derivation of the Equivariant Filter (EqF), an observer design approach based on the principles of the extended Kalman filter applied in equivariant error coordinates.
The approach taken is based on the extended Kalman filter derivation, adapted to continuous-time and the equivariant setting.
In particular, we design a linearised or perturbation Kalman filter \cite{1982_Maybeck_StochasticModels_v2} around a reference trajectory generated by unforced observer dynamics and then translate this reference trajectory to track the filter trajectory by adding a steering input.
The approach allows us to clearly separate the filter dynamics from the re-centering process and consequently identify the role of parallel transport in the covariance reset on a manifold.
The approach is summarised as follows:
\begin{enumerate}
\item Define a reference trajectory and reference error.
\begin{itemize}
  \item The reference trajectory is a solution to the unforced observer dynamics
  \item The reference error is analogous to the equivariant error.
\end{itemize}
\item Linearise the error dynamics around the reference trajectory.
\item Apply a Kalman filter to the linearised error dynamics.
    \begin{itemize}
    \item Implement the Riccati equation in linearised error coordinates.
    \item Instead of directly implementing the linearised error dynamics, define observer dynamics using the full nonlinear internal model of the system along with the correction term from the Kalman filter innovation lifted to the observer state space such that the resulting error trajectory implicitly implements the filter error dynamics.
    \end{itemize}
\item\label{item:EKF_method_steering} Modify the reference trajectory to include a steering term.
    \begin{itemize}
      \item The steering term holds the filter estimate of the error to zero.  That is, to steer the reference trajectory to follow the observer trajectory exactly.
    \end{itemize}

\item Update the filter equations for the new reference trajectory.
\end{enumerate}

\subsection{Define a reference trajectory and reference error}\label{sub:EqF_reference_traj}
The equivariant filter is based on linearised error dynamics computed around a known reference trajectory of the system.
The first step is to choose a reference trajectory $\check{X}(t) \in \grpG$ in the symmetry group and not a trajectory on the system state space as would be the case for a classical EKF.
We choose $\check{X}$ as a solution of the lifted system \eqref{eq:lifted_system}
\begin{align}
\ddt \check{X} = \check{X} \Lambda(\phi_{\mr{\xi}}(\check{X}), u), \quad \check{X}(0) = \Id
\label{eq:lintraj}
\end{align}
This choice is made in order to define the \emph{equivariant reference error}:
\begin{align}
\check{e} = \phi_{\check{X}^{-1}}(\xi).
\label{eq:lin_error}
\end{align}
This error is analogous to the equivariant error \eqref{eq:e} except that instead of measuring the error from state to observer, it measures the error from state to reference trajectory $\check{X} \in \grpG$.
Since the reference trajectory has initial condition $\check{X}(0) = \Id$ we choose the origin $\mr{\xi}$ to be the best \emph{a-priori} guess of the true value of the initial state\footnote{
This assumption can be generalised to arbitrary $\mr{\xi}$ but the choice made above simplifies the following discussion.
}.
As a consequence the initial reference error $\check{e}(0) \sim \GP(\mr{\xi},\Sigma_0)$ is distributed around the origin according to some initial prior distribution.

The analogy with classical Extended Kalman Filtering (EKF) with state  $\xi \in \R^m$ is instructive.
The reference trajectory $\check{\xi} \in \R^m$ is chosen to be a solution of the system model $\dot{\check{\xi}} = f(\check{\xi},u)$,  with $\check{\xi}(0) \approx \xi(0)$ chosen in the neighbourhood of the state.
The linear structure of $\R^m$ leads to a reference error $\tilde{\xi} = \xi - \check{\xi}$.
The error is analysed in a neighbourhood of the origin $0 \in \R^m$.
That is the error is small when $\tilde{\xi}$ is close to the origin $0$.

Analogous to the equivariant output, we define an equivariant reference output
\[
\check{d} = \rho_{\check{X}^{-1}} (y).
\]
One has $\check{d} = h (\check{e})$ analogous to \eqref{eq:d_of_e}.

Define the \emph{reference input} signal $\check{u} := \psi_{\check{X}^{-1}}(u)$ analogous to the origin input discussed earlier \eqref{eq:origin_velocity}.
The full nonlinear dynamics for the equivariant reference error $\check{e}$ around the reference trajectory $\check{X}$ are analogous to the error dynamics \eqref{eq:error_dynamics_equivariant}
\begin{subequations}\label{eq:linerror-system}
\begin{align}
\ddt \check{e} & = \tD \phi_{\check{e}} \left(\Lambda(\check{e},\check{u}) -  \Lambda(\mr{\xi},\check{u}) \right) \label{eq:linerror-system_e}\\
\check{d} & = h( \check{e} ).   \label{eq:linerror-system_d}
\end{align}
\end{subequations}

The goal is to generate an estimate $\hat{e}$ for the actual reference error $\check{e}$.
Given such an estimate, the associated state estimate is obtained by applying the transformation associated with the reference trajectory
\begin{align}
\hat{\xi} = \phi_{\check{X}}(\hat{e}).
\label{eq:hat_xi_hat_e}
\end{align}
In particular, if $\hat{e} = \check{e}$ then $\hat{\xi} = \phi_{\check{X}}(\phi_{\check{X}^{-1}}(\xi)) = \xi$ as required.
Note that the filter problem is posed in error coordinates $\check{e}$ around $\mr{\xi}$ but that the actual estimate of $\hat{e}$ does not necessarily converge to $\mr{\xi}$ in this formulation since $\check{e}$ does not converge to $\mr{\xi}$.
The $\hat{e}$ and $\check{e}$ trajectories shown in Figure \ref{fig:Parallel_transport} provides a visualisation of the formulation.

\subsection{Linearise the error dynamics around the reference trajectory} The equivariant filter (EqF) is based on linearisation of the $\check{e}$ error dynamics.
To linearise a system on a manifold it is necessary to work in a set of local coordinates.
For an equivariant filter, only a single chart of local coordinates on $\calM$ around $\mr{\xi}$ and single chart of coordinates on $\calN$ around $\mr{y} = h(\mr{\xi})$ are required.

Let $\epsilon : \calM \to \R^m$ and $\delta : \calN \to \R^n$ be such local coordinates and we will assume that $\check{e}$ and $\check{d}$ remain in the domain of definition of the local coordinates for all time.
This is to be expected since the reference trajectory $\check{X}$ is the lifted dynamics and will project down to system trajectories.
Expressing the equivariant error system \eqref{eq:linerror-system} in local coordinates and linearising around the origin $\mr{\xi} \in \calM$, and $\mr{y} \in \calN$ yields a time-varying linear system \cite{vanGoor2020_EqF.tac}
\begin{subequations} \label{eq:check_linearisation}
\begin{align}
\td \check{\epsilon} & = A_t \check{\epsilon} \dt + B_t \td \vmu \label{eq:linloccheck_epsilon} \\
\td \check{\delta} & = C_t \check{\epsilon} \dt + D_t \td \ynu \label{eq:linloccheck_delta}
\end{align}
\end{subequations}
where we add noise as Wiener processes $\tD \vmu \sim \WP(0,M_t)$ and $\tD \ynu \sim \WP(0, N_t)$ for positive definite time-varying covariance $M_t, N_t > 0$.
We circumvent the difficulty of understanding noise processes on non-linear manifolds by modelling the noise in the linearisation where classical theory holds.
The linear noise models introduced in the linearisation will correspond to some nonlinear noise model on the full nonlinear system.
A full understanding of noise in equivariant filtering is still an active topic of research.

It is worth taking a moment to discuss the linearised model \eqref{eq:check_linearisation}.
Recall the analogy to the classical EKF where $\xi, \check{\xi} \in \R^m$ and the error is $\tilde{\xi} = \xi - \check{\xi}$.
The EKF error is linearised around $\tilde{\xi} = 0$.
One has
\begin{align*}
\ddt \tilde{\xi} & = \dot{\xi} - \dot{\check{\xi}}
= f(\xi,u) - f(\check{\xi},u)
=  f(\check{\xi} + \tilde{\xi} ,u) - f(\check{\xi},u)
\approx  f(\check{\xi},u) +Df_{u}(\check{\xi}) \tilde{\xi} - f(\check{\xi},u) \\
& = Df_{u}(\check{\xi}) \tilde{\xi}
\end{align*}
That is, linearising the error around $\tilde{\xi} = 0$ is the same is linearising the system function around $\xi = \check{\xi}$.
This leads to direct dependence of the linearisation on the reference trajectory $\check{\xi}$.
In contrast, the error dynamics $\check{e}$ are linearised around the fixed point $\mr{\xi}$ and are independent of the reference trajectory $\check{X}$ except through the transformed input $\check{u} = \psi_{\check{X}^{-1}}(u)$.
Similarly, the $B_t$ and $D_t$ matrices are only dependent on the reference trajectory through the input action, while the output matrix $C_t = \mr{C}$ matrix is constant \cite{vanGoor2020_EqF.tac} since the equivariant innovation is independent of the input.
The regularity of the matrices in the linearisation are a direct consequence of the equivariance of the underlying system  and lead to improved filter performance  \cite{vanGoor2020_EqF.tac}.

\subsection{Apply a Kalman filter to the linearised error dynamics}
From \eqref{eq:check_linearisation} a standard Kalman filter with nonlinear update can be used to filter for the best estimate $\hat{e}$ of $\check{e}$ \cite{Bucy2005_FilteringStochasticProcesses}.
In error coordinates, this filter can be written \cite{vanGoor2020_EqF.cdc,vanGoor2020_EqF.tac}
\begin{subequations}
\label{eq:EKF}
\begin{align}
\ddt \hat{e} & =
\tD \phi_{\hat{e}} \left(\Lambda(\hat{e},\check{u}) -  \Lambda(\mr{\xi},\check{u}) \right) - \tD \epsilon^{-1} \Sigma C_t^\top N_t^{-1} (\delta - \hat{\delta}), \quad \hat{e}(0) = \mr{\xi}
\label{eq:EKF_hat_e}
 \\
\ddt \Sigma & = A_t \Sigma + \Sigma A_t^\top - \Sigma C_t N_t^{-1} C_t^\top \Sigma + B_t^\top M_t B_t, \quad \Sigma(0) = \Sigma_0,  \label{eq:EKF_Sigma}
\end{align}
\end{subequations}
for $\Sigma_0 \in \PD(m)$ the covariance of the prior for the reference $\check{e}(0)$ discussed in Section \ref{sub:EqF_reference_traj}.

The full nonlinear equivariant error dynamics \eqref{eq:EKF_hat_e} are used with the innovation drawn from the linear filter equation as is the usual case in an extended Kalman filter \cite{Bucy2005_FilteringStochasticProcesses}.
The stochastic interpretation of this filter is that the equivariant error $\check{e} \sim \GP(\hat{e},\Sigma)$ is distributed around the estimate $\hat{e}$ with variance $\Sigma$.
The filter is initialised with an estimate of uncertainty for the initial equivariant error $\check{e}(0) \sim \GP(\mr{\xi},\Sigma_0)$.

The filter dynamics \eqref{eq:EKF_hat_e} are a good start, however, they must be related to the observer \eqref{eq:observer} in order to implement a filter.
In the case of equivariant filtering, this involves lifting the error dynamics \eqref{eq:EKF_hat_e} to observer dynamics
\begin{align}
\ddt \hat{X} & = \tD L_{\hat{X}} \Lambda(\phi_{\mr{\xi}} (\hat{X}),\psi_{\hat{X}} (\check{u}))
+ \tD \phi_{\mr{\xi}}^\dag \left( \tD \epsilon^{-1} \Sigma C_t^\top N_t^{-1} (\delta - \hat{\delta}) \right) \hat{X},
\quad \hat{X}(0) = \Id
\label{eq:hatX_observer}
\end{align}
where $\tD \phi_{\mr{\xi}}^\dag : \tT_{\mr{\xi}} \calM \to \gothg$ is a pseudo-inverse of $\tD \phi_{\mr{\xi}}$.
The particular choice of pseudo-inverse is arbitrary and does not effect the performance of the filter.
It is straightforward to verify that $\hat{e}(t) = \tD \phi_{\mr{\xi}} (\hat{X}(t))$ is a solution of \eqref{eq:EKF_hat_e}.
It follows that there is no requirement to implement \eqref{eq:EKF_hat_e} to solve the filtering problem, rather one implements the observer equation  \eqref{eq:hatX_observer} along with the Riccati equation \eqref{eq:EKF_Sigma}.
The EKF filter equation \eqref{eq:EKF_hat_e} is implicit in the equivariant structure of the problem.
The associated state estimate is computed by
\[
\hat{\xi} = \phi_{\check{X}} (\hat{e}) = \phi_{\check{X}} (\phi_{\hat{X}^{-1}} (\mr{\xi})).
\]

\subsection{Modify the reference trajectory to include a steering term}

To this point we have derived a linearised or perturbation Kalman filter
\cite{1982_Maybeck_StochasticModels_v2}, that is, applied the linear Kalman filter to a linearised version of the error dynamics and then lifted these filter dynamics to an observer on the group.
The resulting observer is well defined and provides a stochastic estimate of the \textit{a-positori} distribution of the error $\check{e} \sim \GP(\hat{e},\Sigma) = \GP(\phi_{\mr{\xi}}(\hat{X}),\Sigma)$.
However, the construction relies on a reference trajectory $\check{X}$ that must be chosen such that $\check{e}$ remains close to $\mr{\xi}$.
Choosing initial conditions sufficiently close, then the lifted dynamics of the reference trajectory project to a solution for which the equivariant error will only drift slowly.
Eventually, since the two trajectories are not synchronous \cite{RM_2010_Lageman.TAC}, the equivariant error will drift away from $\mr{\xi}$ compromising the principles of the linearisation on which the filter is based.
This issue is overcome by steering the reference trajectory to keep the error close to $\mr{\xi}$.

Consider the modified reference trajectory dynamics
\begin{align}
\ddt \check{X} = \check{X} \Lambda(\phi_{\mr{\xi}}(\check{X}), u) - \Delta_t \check{X}, \quad \check{X}(0) = \Id.
\label{eq:lintraj_mod}
\end{align}
Where $\Delta_t \in \gothg$ is a steering term.
With this choice then the equivariant error dynamics \eqref{eq:linerror-system_e} become
\begin{align*}
\ddt \check{e} & = \tD \phi_{\check{e}} \left(\Lambda(\check{e},\check{u}) -  \Lambda(\mr{\xi},\check{u}) \right) + \tD \phi_{\check{e}} \Delta_t
\end{align*}
and in consequence the filter dynamics, which are a copy of the error dynamics plus the correction term, become
\begin{align}
\ddt \hat{e} & =
\tD \phi_{\hat{e}} \left(\Lambda(\hat{e},\check{u}) -  \Lambda(\mr{\xi},\check{u}) \right) + \tD \phi_{\hat{e}} \Delta_t
+ \tD \epsilon^{-1} \Sigma C_t^\top N_t^{-1} (\delta - \hat{\delta}), \quad \hat{e}(0) = \mr{\xi}.
\label{eq:EKF_hate_steered}
\end{align}
From here it is natural to choose
\begin{align}
\Delta_t := - \tD \phi_{\hat{e}}^\dag \tD \epsilon^{-1} \Sigma C_t^\top N_t^{-1} (\delta - \hat{\delta}).
\label{eq:Delta}
\end{align}
Noting that $\hat{e}(0) = \mr{\xi}$, it is straightforward to verify that $\hat{e}(t) = \mr{\xi}$ is a solution of \eqref{eq:EKF_hate_steered} for all time.
That is, the steering term $\Delta_t$ is chosen to fix the estimate $\hat{e}\equiv \mr{\xi}$ constant at the origin.
Since the estimate $\hat{e}$ was designed to converge to $\check{e}$ then it follows that $\check{e}$ should converge to $\mr{\xi}$.

An important consequence of this choice is that the modified reference trajectory \eqref{eq:lintraj_mod} is also a solution of the observer \eqref{eq:hatX_observer}.
This follows since the initial conditions match and along a trajectory of $\check{X} = \hat{X}$ then $\check{u} = \mr{u}$ and the defining equations become identical.
Thus, the role of the reference trajectory is subsumed by the observer and \eqref{eq:lintraj_mod} can be discarded.
Moreover, the equivariant reference error is now
\[
\check{e} = \phi_{\check{X}^{-1}} (\xi) = \phi_{\hat{X}^{-1}} (\xi) = e;
\]
the equivariant error \eqref{eq:e} introduced in Section \ref{sec:symmetry}.
As such, the notation for $\check{e}$ can also be discarded.
Note that steering the reference trajectory to track the observer trajectory can only be undertaken once the filter estimate is available.
The reference trajectory must be defined first while the linearisation of the error dynamics is computed around a reference trajectory that does not include the steering term.
In a classical extended Kalman filter, this process is a re-centering or reset process done at each step of the algorithm.
In continuous time, it is a steering term added to the unforced internal model that corresponds to the filter correction term.
Thus, the correction term is doing two things, firstly implementing the correction term associated with the linearised Kalman filter algorithms, and secondly steering the reference trajectory to track the filter estimate.

\subsection{Update the filter equations for the new reference trajectory}

The final step in the extended Kalman filter derivation is to combine
\eqref{eq:hatX_observer} and \eqref{eq:EKF_Sigma} (with $\check{u}$ replaced by $\mr{u}$).
The observer trajectory $\hat{X}$ is the reference trajectory, the observer estimate is $\mr{\xi}$ for all time and the equivariant error $e \to \mr{\xi}$ at least as long as the filter converges.
However, the derivation of the filter used a linearisation associated with a reference trajectory that was a solution of the system \eqref{eq:lintraj}, and steering the trajectory breaks this assumption.
In fact, the steering term corresponds to translating the filter solution in the error coordinates in order to keep $\hat{e}$ centred on the origin $\mr{\xi}$.
For the error dynamics themselves, this is just a change of coordinates and does not change the observer dynamics of the filter.
But the filter estimate is not just the observer state $\hat{e}$, it is the full information state estimate $\check{e} \sim \GP(\hat{e},\Sigma)$ and translating this solution involves parallel transport of the covariance $\Sigma$.

\begin{figure}
  \centering
  \includegraphics[scale=0.56]{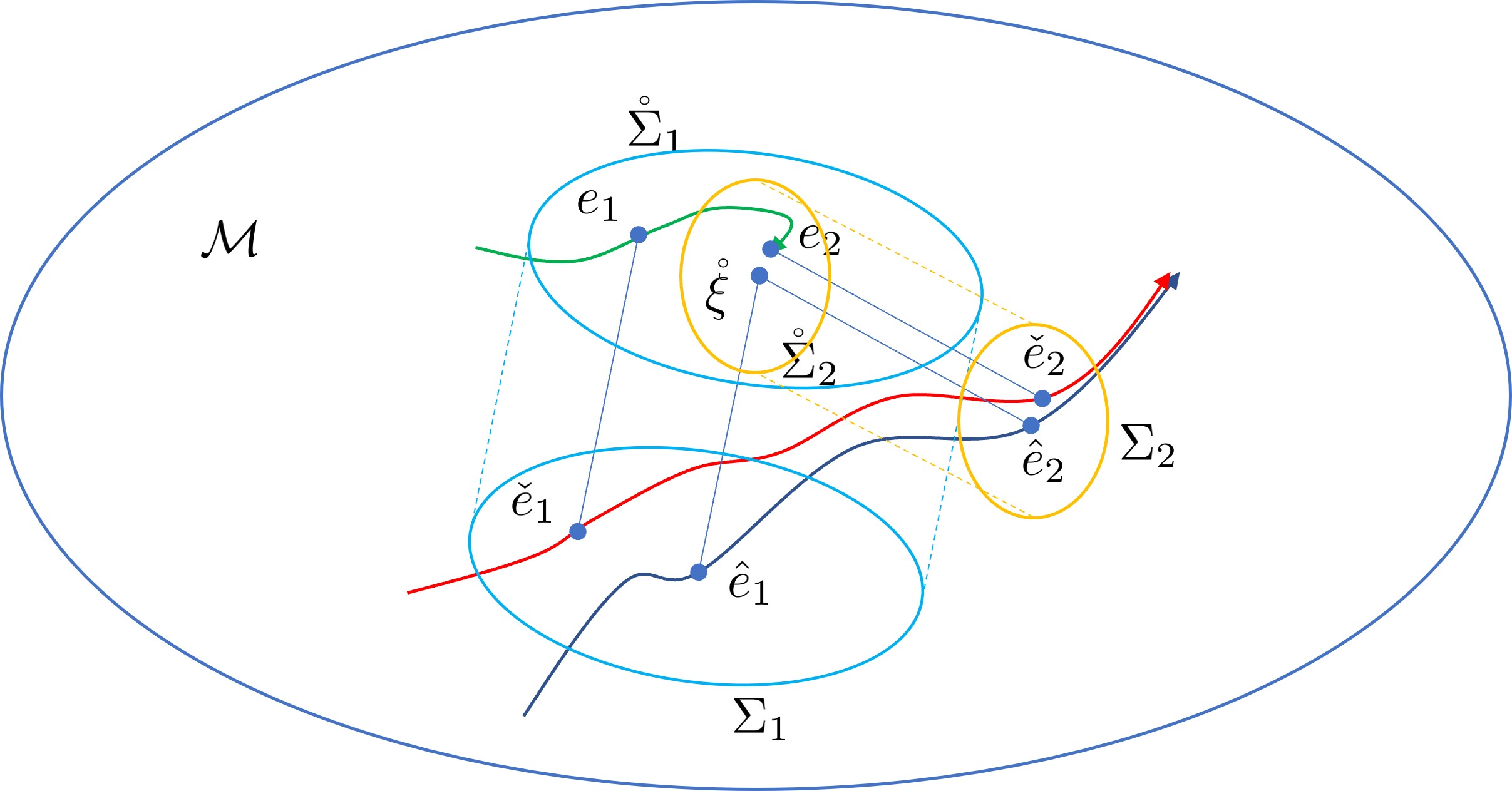}
  \caption{The evolution of error variables and covariances shown at two times with subscripts one and two.
  As time passes, the trajectory of $\hat{e}$ converges to $\check{e}$ and the covariance estimate $\Sigma$ contracts.
  The information state for the reference error $\check{e}_i \sim \GP(\hat{e}_i,\Sigma_i)$ transforms to an information state for the equivariant observer error $e_i \sim \GP(\mr{\xi},\mr{\Sigma}_i)$.    The covariance $\mr{\Sigma}$ is the parallel transport of the covariance $\Sigma$.
  }
  \label{fig:Parallel_transport}
\end{figure}

Figure \ref{fig:Parallel_transport} shows a figure where the unmodified filter solution generates an estimate $\GP(\hat{e},\Sigma)$ that converges to the true equivariant error state $\check{e}$.
Adding the steering term to the reference trajectory is analogous to translating the estimate $\hat{e}$ to the origin $\mr{\xi}$, with a corresponding translation of the reference error $\check{e}$ to the equivariant error $e$.
The covariance $\Sigma$ is translated to the corresponding covariance $\mr{\Sigma}$ around $\mr{\xi}$.
The transformed filter state can be interpreted as the information state for the equivariant error $e \sim \GP(\mr{\xi},\mr{\Sigma})$.
If the manifold is flat then the parallel transport is trivial and $\mr{\Sigma} = \Sigma$, however, if the manifold has curvature this additional parallel transport should be modeled as a curvature modification to the Riccati equation.

Recall that parallel transport $\PT$ on a manifold is defined along curves on the space.
If $\gamma : \R \to \calM$ is a smooth curve then parallel transport is an operator $\PT_{\gamma} : \tT_{\gamma(0)} \calM \to \tT_{\gamma(t)} \calM$ that maps tangent vectors $g \in \tT_{\gamma(0)} \calM$ to their parallel transport $\PT_{\gamma(t)} g \in \tT_{\gamma(t)} \calM$ for $t \geq 0$.
Covariance $\Sigma \in \PD(m)$ can be thought of as a (0,2)-tensor that accepts two tangent vectors (written in local coordinates) and returns a scalar.
The parallel transport of $\Sigma$ over the curve $\gamma$ is given by
$ \mr{\Sigma} = \PT_{\gamma} \Sigma \PT_{\gamma}^{\top}$ (in local coordinates).

Since the steering process is an infinitesimal modification of the continuous time reference trajectory, the correction to $\Sigma$ does not need to be computed explicitly as a parallel transformation.
Rather, the steering input can be applied to the covariance evolution as an instantaneous curvature correction.
The infinitesimal parallel transport of a vector $g \in \tT_{\gamma(0)} \calM$ due to an infinitesimal displacement $\dot{\gamma}(0)$ is
\begin{align*}
 \left. \ddt \right|_{t=0} \PT_{\gamma(t)} g
 =: \Gamma_{\dot{\gamma}(0)} g,
\end{align*}
where $\Gamma_{\dot{\gamma}(0)} : \tT_{\gamma(0)}\calM \to \tT_{\gamma(0)}\calM$ is the connection function for the manifold $\calM$ evaluated at $\gamma(0)$ in direction $\dot{\gamma}(0)$.
Consequently, the infinitesimal modification of $\Sigma$ is given by
\begin{align*}
 \left. \ddt \right|_{t=0} \mr{\Sigma}
  &= \ddt |_{t=0} \PT_{\gamma(t)} \Sigma  \PT_{\gamma(t)}^\top, \\
  &= \Gamma_{\gamma(0)} \Sigma + \Sigma \Gamma_{\gamma(0)}^\top.
\end{align*}
In the final line we write the connection function for a known infinitesimal translation  $\Gamma_{\dot{\gamma}(0)} \in \R^{m \times m}$ as a matrix in the same local coordinates as those used for $\Sigma$.

Now, in our case the steering input applies an infinitesimal translation $\dot{\gamma}(0) = - \tD \phi_{\mr{\xi}} \Delta_t$ to the trajectory $\hat{e}(t)$.
This point is somewhat subtle as in fact this term is holding $\hat{e}(t) \equiv \mr{\xi}$, nevertheless, the infinitesimal translation is transporting the covariance that otherwise would have been moving along with the reference system trajectory.
Substituting the known infinitesimal translation due to the steering input into the parallel transport equations one obtains the required modification term to the Riccati equation.

Thus, following the methodology outlined at the start of Section \ref{sec:design} for an equivariant system, we propose the Equivariant Filter (EqF) with curvature modification
\begin{subequations}\label{e:EqF}
\begin{align}
\Delta_t & := \tD \phi_{\mr{\xi}}^\dag  \tD \epsilon^{-1} \Sigma C_t^\top N_t^{-1} (\delta - \hat{\delta}),  \label{eq:EqF_Delta} \\
\ddt \hat{X} & = \hat{X} \Lambda(\phi_{\mr{\xi}}(\hat{X}), u) -
\Delta_t \hat{X}, \quad \hat{X}(0) = \Id \label{eq:EqF_hatX} \\
\ddt \Sigma & = A_t \Sigma + \Sigma A_t^\top - \Sigma C_t N_t^{-1} C_t^\top \Sigma + B_t^\top M_t B_t
-\Gamma_{ \tD \phi_{\mr{\xi}} \Delta_t} \Sigma
- \Sigma \Gamma_{\tD \phi_{\mr{\xi}} \Delta_t}^{\top},
\quad \Sigma(0) = \Sigma_0 \label{eq:EqF_Sigma}
\end{align}
\end{subequations}
where the ``curvature modification'' $-(\Gamma_{\tD \phi_{\mr{\xi}} \Delta_t} \Sigma  +  \Sigma \Gamma_{\tD \phi_{\mr{\xi}} \Delta_t}^{\top})$ in \eqref{eq:EqF_Sigma} compensates for the distortion of the covariance due to parallel transport on a non-flat manifold.
Note that since $\Delta_t$ depends on $\Sigma$, the curvature modification is a quadratic perturbation to the classical Ricatti equation.

It is interesting to note that the curvature modification \eqref{eq:EqF_Sigma} is driven by the steering term $\tD \phi_{\mr{\xi}} \Delta_t$ and not the system dynamics.
In particular, the nonlinearity associated with the evolution of the system is captured in the local coordinates and is modeled in the linearisation process used to derive the Kalman filter in error coordinates.
It is only the process of re-centering the error using a steering input that incurs the curvature correction.
The re-centering process is fundamental for the EqF where the filter is  continuously translated to centre on the origin, a process that yields significant benefits in the regularity of the linearisation mentioned earlier.

\section{Simulations}\label{sec:simulations}

In this section we provide simulations to demonstrate the qualitative behaviour of the equivariant filter on the two examples discussed in the body of the paper.

In Section \ref{sub:direction} we consider the direction kinematics \eqref{eq:direction_kinematics} and compare an EqF using normal coordinates inherited from the symmetry with an EKF using stereographic coordinates.
In Sections \ref{sub:Galilean} and \ref{sub:polar} we consider second order linear kinematics with range and bearing measurements.
In Section \ref{sub:Galilean} we simulate two filters both using the Galilean symmetry.
In the first Galilean filter we algebraically reconstruct a position measurement from the actual range and bearing measurements and apply a linear Kalman filter.
In the second we use the classical extended Kalman filter (EKF) in the natural Euclidean coordinates on $\R^3 \times \R^3$ by linearising the bearing and range measurement functions.
Section \ref{sub:polar} uses the polar symmetry developed in Section \ref{sec:outputs}.
We implement the full EqF in this symmetry for the range and bearing measurements, taking care to use local coordinates inherited from the symmetry of the system.
We also simulate the same filter, in the same local coordinates, but without the curvature modification term in the Riccati equation \eqref{eq:EqF_Sigma} for comparison.

All simulations were done by implementing the continuous-time system equations using Euler integration at a constant time-discretization of $\dt = 0.02$s.
Noise was simulated as Gaussian processes added directly to the variables at each time step.

\subsection{Direction kinematics: Rotational symmetry} \label{sub:direction}

For simplicity we assume the true system is rotating with a constant body velocity $\Omega = (0, 0.5, -0.2)$rad/s.
The input for the systems is a velocity measurement $v = \Omega + \vmu$  corrupted with Gaussian noise $\vmu \sim \GP(0, 0.01)$ in rad/s.
The initial value of the true direction $\eta \in \Sph^2$ was chosen to be
$\eta = (\eb_3 + \mu)/|\eb_3 + \mu|$, where $\mu \sim \GP(0, 10.0)$.
The measurement $y = (\eta + \ynu)/|\eta + \ynu|$ was corrupted by Gaussian noise $\ynu \sim \GP(0, 0.1)$.
Figure~\ref{fig:eqf_sphere} plots both the raw state error, in this case the bearing error $\theta = \vert \arccos(\hat{\eta}^\top \eta) \vert $, and the filter energy.
The filter energy refers to the quantity $\Lyap = \varepsilon^\top \Sigma^{-1} \varepsilon$ that describes the log-likelihood of the estimate with respect to the filter's own estimate of the covariance $\Sigma$ of the error estimate $\varepsilon$.
Low filter energy indicates internal consistency of the filter, and is a desirable feature alongside error in the state estimate.
Figure~\ref{fig:eqf_sphere} uses a $\log_{10}$ scale on the y-axis to show the convergence properties of the errors more clearly.

The first filter we implement is a classical EKF.
For this filter we use the well-known stereographic projection for local coordinates, centred on the filter estimate $\hat{\eta}$.
That is, we adapt the local coordinates to track with the filter to minimize the linearisation error associated with poor local coordinate conditioning.
The second filter is the EqF where we use the equivariant error around the origin  $\eb_3 \in \Sph^2$.
For this filter we use the so called \emph{normal coordinates} on $\calM$, obtained by projecting exponentials from the Lie-group to the manifold via the group action.
Specifically, one sets $\varepsilon(y) = \omega$, where $\omega \in \R^2$ is the unique element such that $\exp((\omega,0)^\times) \eb_3 = y$.
These coordinates most closely capture the equivariance of the system and can be thought of as the natural generalisation of the log-linear coordinates that play a critical role in the analysis of group affine systems on Lie-groups \cite{2017_Barrau_tac}.
Since the linearisation point is fixed, only the single local coordinate chart is required for the EqF.
The sphere is a reductive homogeneous space of $\SO(3)$ and hence has a unique symmetric connection\footnote{
The projection of the \emph{Cartan-Schouten (0)} or \emph{(first) canonical} connection on the Lie-group.
The canonical connection on the Lie group is the unique \emph{symmetric} connection for which geodesics are the 1-parameter subgroups (exponentials).
The normal coordinates (projected exponential) on the homogeneous space are geodesics of the induced connection.
} that is compatible with the symmetry action \cite{1963_Kobayashi_foundations}.
This connection corresponds to the normal Riemannian connection on $\Sph^2$ and as a consequence the connection coefficients $\Gamma_{ij}^k$ are zero in the normal coordinates at $\eb_3$.
It follows that the curvature modification terms $\Gamma_{\tD \phi_{\mr{\xi}} \Delta_t}$ in \eqref{eq:EqF_Sigma} are uniformly zero at the origin $\mr{\eta}$.
It is important to note that this property is closely coupled to the particular choice of local coordinates and these terms are not zero in the stereographic coordinates.

Figure \ref{fig:eqf_sphere} shows the performance of both filters.
The EqF with normal coordinates clearly outperforms the EKF with stereographic coordinates as shown.
The main reason for the performance advantage lies in the choice of normal  coordinates as the local coordinate chart that minimizes linearisation error.
This is particularly clear early in the trajectory when the correct scaling of local coordinates significantly reduces the transient of the filter.

\subsection{Second order kinematics: Galilean symmetry}\label{sub:Galilean}

The trajectory for the second order kinematics was an oscillatory trajectory with acceleration $a = (0, \cos(5 t), 0)$.
The initial state was chosen $p(0) = \phi_{\exp(\mu^\wedge)}(0, 0, 50)$ and $v = 0$ where the variation in position $|\mu| \sim \GP(0,0.25)$ is generated by a homogeneous Gaussian on the group projected to normal coordinates on the state manifold.
The acceleration measurement $a = a  + \vmu$ is was corrupted by Gaussian noise $\vmu \sim \GP(0,0.0025)$ in m/s$^2$.
The bearing and range  measurements \eqref{eq:y1y2}  were corrupted by Gaussian noise $\nu_1 \sim \GP(0,4)$ in degrees and $\nu_2 \sim \GP(0,4)$ in metres respectively.
Figure~\ref{fig:eqfolar} shows position error $|p - \hat{p}|$
and velocity error $|v - \hat{v}|$ to show the tracking error.
We also show the filter energy analogous to Figure~\ref{fig:eqf_sphere}.
All plots are in $\log_{10}$ scale to better show the convergence properties of the error signals.

The Galilean symmetry preserves the linear structure of the state space $\R^3 \times \R^3$.
The error $e = \phi_{(\hat{\alpha},\hat{\beta})^{-1}} (p,v) = (p -\alpha, v - \beta)$ and the unforced error dynamics are a linear time invariant system
\begin{align}
\dot{e} & = \left( \dot{\tilde{p}}, \dot{\tilde{v}} \right) = (\tilde{v} , 0) \label{eq:linear_error_kin}
\end{align}
The measurements $y_1$ and $y_2$ are bearing and range \eqref{eq:y1y2} and it is possible to reconstruct an algebraic estimate of a position measurement
\begin{align*}
  y = y_1 y_2 = p + \ynu,
\end{align*}
with noise model $\ynu \sim \GP(0, 4 y_2^2 (I_3 - y_1 y_1^\top) + 4 y_1 y_1^\top)$ estimated from the noise characteristics of $y_1$ and $y_2$.
With full position measurements $y = p$ then the innovation is a linear function of the error state $\tilde{y} = \tilde{p}$ and a linear Kalman filter applies directly to this system.

Applying the principles of the extended Kalman filter to the full nonlinear model, the nonlinear measurements $y_1$ and $y_2$ are linearised.
The linearisation of the associated equivariant innovations are
\begin{align*}
\tilde{y}_1= \left( I_3 - \frac{\mr{p}\mr{p}^\top}{|\mr{p}|^2} \right)  \tilde{p}, \quad\quad\quad
\tilde{y}_2 = \frac{\mr{p}^\top}{|\mr{p}|} \tilde{p}.
\end{align*}
The extended Kalman filter uses these measurement functions along with the linear error kinematics \eqref{eq:linear_error_kin}.

Figure~\ref{fig:eqfolar} shows qualitative results for these two filters in
solid red and dashed blue.
The linear Kalman filter is actually highly effective as long as the noise in the measurements is low.
However, it is very prone to instability for increasing measurement noise.
The particular noise characteristics we have chosen are on the limit of the envelope in which the linear Kalman filter is effective and the trajectory  demonstrates incipient instability in some of the sharp jumps in the filter response.
For lower noise level the performance of the linear Kalman filter is similar to that of the EqF filters and indeed, if the measurement $y = p$ were truly distributed as a Gaussian variable then the equivariant filter in Galilean symmetry is exactly the linear Kalman filter.
For higher noise levels the linear Kalman filter diverges.
In contrast, the EKF is robust and stable, however, its performance is clearly inferior to the other filters.
This is due to the linearisation error incurred by linearising the output functions around the incorrect state estimates.

\subsection{Second order kinematics: Polar symmetry} \label{sub:polar}

The equivariant filters for second order linear kinematics with range and bearing measurements were implemented for the polar symmetry.
The same system trajectories and parameters were used with identical noise sequences.

For the polar symmetry the lift function is
\begin{align*}
\Lambda((p,v),a) & :=
\left(
\frac{(p \times v)^\times}{\vert p \vert^2},
\frac{p \cdot v}{\vert p \vert^2},
\left[
\frac{(p \times v) \times v + (p \cdot v) v }{\vert p \vert^2} - a
\right]
\right)
\in (\so(3) \times \mathfrak{mr}(1)) \ltimes \gothr^3
\end{align*}
that maps into the Lie-algebra of $(\SO(3) \times \MR(1)) \ltimes \grpR^3$.
The lifted system is
\begin{align*}
\dot{R} & = -R \frac{(p \times v)^\times}{\vert p \vert^2}, &
\dot{r} & = -r \frac{p \cdot v}{\vert p \vert^2}, &
\dot{\beta} & = r R \left(\frac{(p \times v) \times v + (p \cdot v) v }{\vert p \vert^2} - a \right),
\end{align*}
where $(p, v) = \phi((R,r,\beta), (\mr{p}, \mr{v}))$.
The error dynamics are complicated and we omit them to save space.
The origin element\footnote{
The group action for the polar symmetry is not actually transitive on the whole of $\R^3 \times \R^3$; the set $\{0\} \times \R^3$ and $\R^3/\{0\} \times \R^3$ are disjoint orbits of the action.
The origin must be chosen in the orbit that where the trajectory will evolve.
This is not a true restriction since the measurements are not defined on the exception set either.
}
chosen was $((0,0,50),(0,0,0)) \in \R^3 \times \R^3$.
The coordinates used are the normal coordinates obtained by projecting the normal exponential coordinates from the Lie-group down onto the state space around the origin.
Only a single coordinate chart is required.

The range and bearing measurements are equivariant with respect to the polar symmetry.
As a result, the measurements can be written using a subset of the same exponential coordinates as the state elements.
In this particular case, this leads to very simple and state-independent linearisations of the output:
\begin{align*}
  \delta_1 = \begin{pmatrix}
    I_2 & 0_{2,4}
  \end{pmatrix} \varepsilon,\quad\quad\quad
  \delta_2 = \begin{pmatrix}
    0_{1,2} & 1 & 0_{1,3}
  \end{pmatrix} \varepsilon.
\end{align*}
The regularity of the output linearisation is one of the key properties of the approach that leads to the performance of the EqF.

The homogeneous space is reductive (the stabilizer is $S^1$) and we will use the canonical symmetric invariant connection induced on the state manifold by the canonical invariant connection on the Lie-group.
However, unlike the direction kinematics on the sphere, the resulting connection is not a Riemannian connection and the connection coefficients are not zero in the normal coordinates.
Using exponential coordinates of the Lie group to describe the manifold about the origin $(\mr{p}, \mr{v})$, the connection coefficients are\footnote{
This choice, the normal connection, is the unique symmetric connection for which projected exponentials are geodesics on the manifold.
A full discussion of affine invariant connections on homogeneous spaces is beyond the scope of the present paper.
}
\begin{align*}
  \Gamma_\Delta = \frac{1}{2} \ad_{\Delta} \vert_\gothm,
\end{align*}
for any $\Delta \in \gothm$, where the tangent space about $(\mr{p},\mr{v})$ is identified with a subspace $\gothm$ of the Lie algebra, $\tT_{(\mr{p}, \mr{v})} \calM \simeq \gothm \subset \gothg$ invariant under $\Ad_{\stab_\phi(\mr{\xi})}$.

\begin{figure}
  \includegraphics[width=0.8\linewidth]{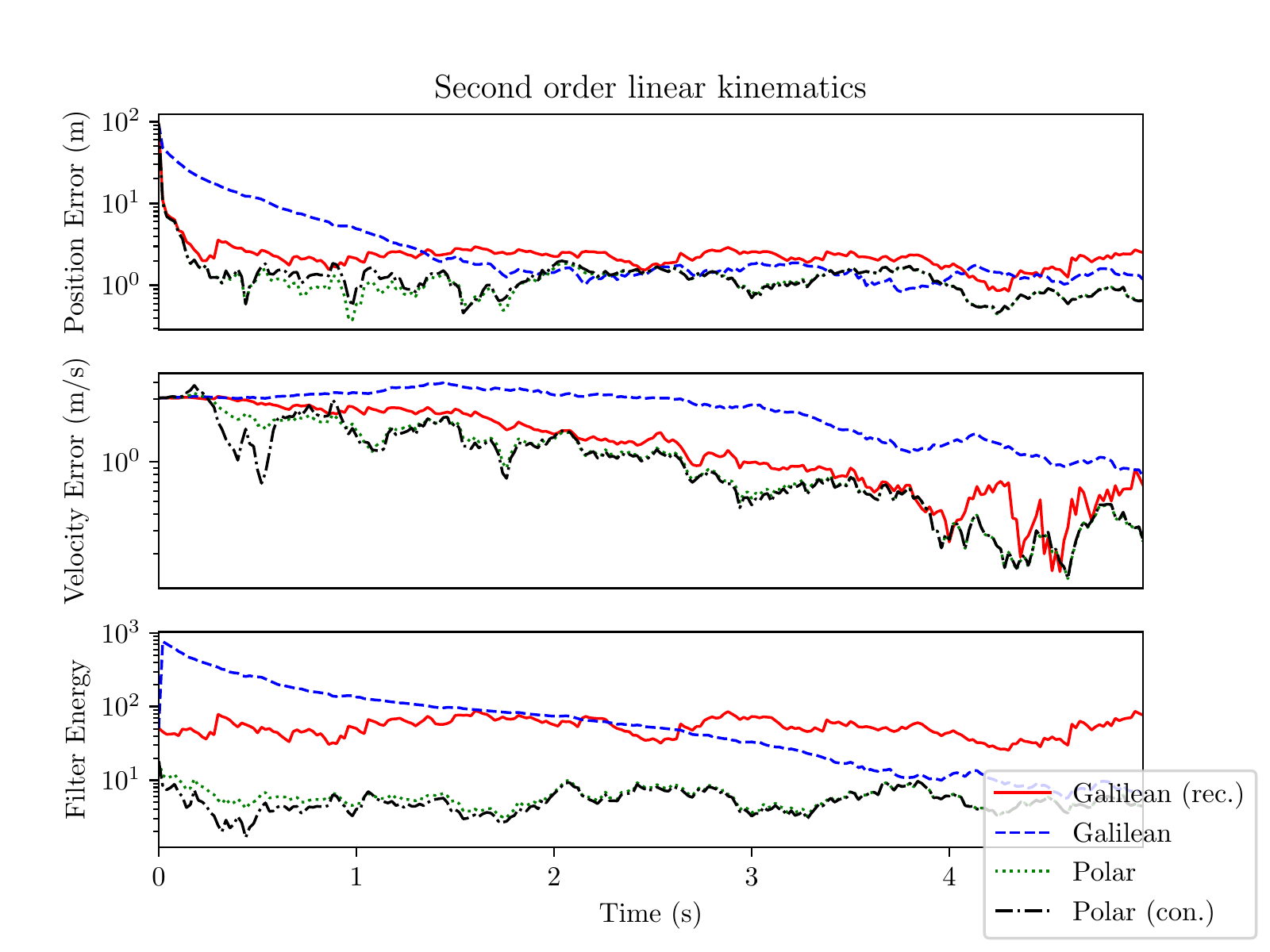}
  \caption{
  Comparison of four different filters for second order kinematics with bearing and range measurement.
  The red solid line shows the linear Kalman filter with reconstructed position measurements.
  The blue dashed lines show the EKF with Galilean symmetry and linearised output.
  The green dotted lines show the EqF with polar symmetry but without curvature modification.
  The black dot-dashed lines show the EqF with polar symmetry and curvature modification.}
  \label{fig:eqfolar}
\end{figure}

Figure \ref{fig:eqfolar} shows the qualitative performance of two filters: the EqF with (black, dash-dot) and without (green, dotted) curvature modification.
It is clear that both these filters outperform the extended Kalman filter and the reconstructed linear Kalman filter.
The bulk of this performance advantage comes from the equivariance of the output and nice form of the equivariant innovation.
This structure minimizes the linearisation error in the filter and leads to significant performance gains.
The performance difference between the two EqF filters is relatively small.
The curvature correction appears to make most difference in the transient of the velocity term.
This is likely due to the fact that these variables are not directly measured and rely on the quality of the covariance estimate to couple the innovation into a state correction.
It is intuitive that the curvature modification will make most difference when $\Delta_t$ is large during the transient, and that the difference will be most visible in states that are not directly measured.
Once the correction term $\Delta_t$ is small, the curvature modification appears to make little difference to the asymptotic performance of the filter, although it is apparent that the filter energy for the full EqF is always superior to that without the curvature modification.
Given the complexity of the curvature correction it is not clear that its benefits are worth the extra effort in derivation and implementation unless either performance is at an absolute premium or the filter is often in transient.

It is clear from the results obtained that the two EqF filters significantly outperform the standard EKF filter design.
As noted above, for low noise levels the linear Kalman filter with reconstructed error performs similarly to the EqF filters, however, it is not robust to increased noise levels as shown in the simulation.

\section*{ACKNOWLEDGMENTS}

This research was supported by the Australian Research Council
through Discovery Grant DP210102607 `` Exploiting the Symmetry of
Spatial Awareness for 21st Century Automation'' and the Franco-Australian International Research Project “Advancing Autonomy for Unmanned Robotic Systems” (IRP ARS).

\newcommand{\etalchar}[1]{$^{#1}$}


\end{document}